\documentclass[12pt]{article}
\PassOptionsToPackage{hyphens}{url}
\usepackage[pagebackref=true,colorlinks]{hyperref}
\usepackage{amsmath,amssymb,amsthm,mathtools,amsrefs,mathrsfs}
\usepackage{pifont}
\usepackage{mathpazo} 
\usepackage{eulervm}
\usepackage{csquotes}
\usepackage{lscape}
\usepackage[table]{xcolor}
\usepackage{tikz}
\usetikzlibrary{arrows.meta}
\usetikzlibrary{decorations.pathmorphing,shapes}
\usetikzlibrary{calc}
\usepackage{comment}
\usepackage{adjustbox}
\usepackage{scalerel,stackengine}
\usepackage{stmaryrd}
\usepackage{fancyhdr}
\usepackage{soul}
\usepackage{dsfont}
\usepackage[normalem]{ulem}
\usepackage{longtable}
\usepackage[bottom]{footmisc}
\usepackage{graphicx}
\usepackage{caption}
\usepackage{subcaption}
\usepackage{pictexwd, dcpic}
\usepackage{float}
\usepackage{enumerate}
\usepackage[all]{xy} 
\usetikzlibrary{circuits.ee.IEC}
\hypersetup{
	colorlinks=true,
    linktoc=all,
    linkcolor=blue,  
    citecolor=blue,
    }

\usepackage{tocbasic}
\DeclareTOCStyleEntry[
  beforeskip=-0.3em plus 1pt,
  pagenumberformat=\textbf
]{tocline}{section}

\makeatletter
\newcommand*{\shifttext}[2]{%
  \settowidth{\@tempdima}{#2}%
  \makebox[\@tempdima]{\hspace*{#1}#2}%
}
\makeatother
\makeatletter
\renewcommand*\env@matrix[1][\arraystretch]{%
  \edef\arraystretch{#1}%
  \hskip -\arraycolsep
  \let\@ifnextchar\new@ifnextchar
  \array{*\c@MaxMatrixCols c}}
\makeatother

\stackMath
\newcommand\reallywidehat[1]{%
\savestack{\tmpbox}{\stretchto{%
  \scaleto{%
    \scalerel*[\widthof{\ensuremath{#1}}]{\kern.1pt\mathchar"0362\kern.1pt}%
    {\rule{0ex}{\textheight}}
  }{\textheight}%
}{2.4ex}}%
\stackon[-6.9pt]{#1}{\tmpbox}%
}
\theoremstyle{plain}
\newtheorem{theorem}[equation]{Theorem}
\newtheorem{lemma}[equation]{Lemma}
\newtheorem{proposition}[equation]{Proposition}
\newtheorem{corollary}[equation]{Corollary}
\theoremstyle{definition}
\newtheorem{definition}[equation]{Definition}
\newtheorem{construction}[equation]{Construction}
\newtheorem{question}[equation]{Question}
\newtheorem{convention}[equation]{Convention}
\newtheorem{problem}[equation]{Problem}
\newtheorem{example}[equation]{Example}
\newtheorem{exercise}[equation]{Exercise}
\newtheorem*{answer}{Answer}
\newtheorem*{solution}{Solution}
\newtheorem{remark}[equation]{Remark}

\newtheorem{notation}[equation]{Notation}
\newtheorem{noterm}[equation]{Notation and Terminology}

\newcommand\define[1]{\emph{\textbf{#1}}}

\numberwithin{equation}{section}

   \let\de=\delta

\newcommand{\be}{\begin{equation}}
\newcommand{\ee}{\end{equation}}
\def\ba{\begin{align}} 
\def\ea{\end{align}}
\newcommand{\bea}{\begin{eqnarray}}
\newcommand{\eea}{\end{eqnarray}}
\newcommand{\bx}{\begin{example}}
\newcommand{\ex}{\end{example}}
\newcommand{\bex}{\begin{exercise}}
\newcommand{\eex}{\end{exercise}}
\newcommand{\ban}{\begin{answer}}
\newcommand{\ean}{\end{answer}}
\newcommand{\bt}{\begin{theorem}}
\newcommand{\et}{\end{theorem}}
\newcommand{\bc}{\begin{corollary}}
\newcommand{\ec}{\end{corollary}}
\newcommand{\blem}{\begin{lemma}}
\newcommand{\elem}{\end{lemma}}
\newcommand{\bp}{\begin{problem}}
\newcommand{\ep}{\end{problem}}
\newcommand{\bn}{\begin{proposition}}
\newcommand{\en}{\end{proposition}}
\newcommand{\bd}{\begin{definition}}
\newcommand{\ed}{\end{definition}}
\newcommand{\bcon}{\begin{construction}}
\newcommand{\econ}{\end{construction}}
\newcommand{\bq}{\begin{question}}
\newcommand{\eq}{\end{question}}
\newcommand{\bprf}{\begin{proof}}
\newcommand{\eprf}{\end{proof}}
\newcommand{\br}{\begin{remark}}
\newcommand{\er}{\end{remark}}
\newcommand{\bs}{\begin{solution}}
\newcommand{\es}{\end{solution}}
\newcommand{\beqs}{\begin{eqnarray}}
\newcommand{\eeqs}{\end{eqnarray}}
\newcommand{\bnt}{\begin{noterm}}
\newcommand{\ent}{\end{noterm}}
\newcommand{\bnot}{\begin{notation}}
\newcommand{\enot}{\end{notation}}

 \let\ov=\overline

\newcommand{\id}{\mathrm{id}}

\newcommand{\mC}{\mathcal{C}}
\newcommand{\mM}{\mathcal{M}}
\newcommand{\mD}{\mathcal{D}}

\newcommand{\ra}{\rightarrow}

\def\R{{{\mathbb R}}}

\def\N{{{\mathbb N}}}

\def\B{{{\mathbb B}}}

\newcommand\aeequals[1]{\underset{\raisebox{0.3ex}[0pt][0pt]{\scriptsize${#1}$}}{=}}

\newcommand{\op}{\mathrm{op}}

\newcommand{\FinStoch}{\mathbf{FinStoch}}
\newcommand{\FinProb}{\mathbf{FinProb}}
\newcommand{\FinPD}{\mathbf{FinPD}}
\newcommand{\FinPS}{\mathbf{FinPS}}

\def\invexcl{\rotatebox[origin=c]{180}{$!$}}
\newcommand{\bloom}{\operatorname{\invexcl}}
\newcommand{\shriek}{\operatorname{!}}

\newcommand{\stoch}{\;\xy0;/r.25pc/:(-3,0)*{}="1";(3,0)*{}="2";{\ar@{~>}"1";"2"|(1.06){\hole}};\endxy\!}
\newcounter{sarrow}
\newcommand\xstoch[1]{%
\stepcounter{sarrow}%
\mathrel{\begin{tikzpicture}[baseline= {( $ (current bounding box.south) + (0,-0.1ex) $ )}]
\node[inner sep=.5ex] (\thesarrow) {\;$\scriptstyle #1$\;};
\path[draw,{<[scale=1.5,width=3,length=2]}-,decorate,
  decoration={snake,amplitude=0.3mm,segment length=2.1mm,pre=lineto,pre length=1pt}] 
    (\thesarrow.south east) -- (\thesarrow.south west);
\end{tikzpicture}}%
}

\newcounter{sqarrow}

\newcommand{\ds}{\displaystyle}
\newcommand{\ben}{\renewcommand{\theenumi}{\alph{enumi}} 
\renewcommand{\labelenumi}{(\theenumi)}\begin{enumerate}}
\newcommand{\een}{\end{enumerate}}

\newcommand\blfootnote[1]{%
  \begingroup
  \renewcommand\thefootnote{}\footnote{#1}%
  \addtocounter{footnote}{-1}%
  \endgroup
}

\newlength\stateheight
\newlength\minimumstatewidth
\tikzset{width/.initial=\minimummorphismwidth}
\tikzset{colour/.initial=white}

\newif\ifblack\pgfkeys{/tikz/black/.is if=black}
\newif\ifwedge\pgfkeys{/tikz/wedge/.is if=wedge}
\newif\ifvflip\pgfkeys{/tikz/vflip/.is if=vflip}
\newif\ifhflip\pgfkeys{/tikz/hflip/.is if=hflip}
\newif\ifhvflip\pgfkeys{/tikz/hvflip/.is if=hvflip}

\pgfdeclarelayer{foreground}
\pgfdeclarelayer{background}
\pgfsetlayers{background,main,foreground}

\def\thickness{0.4pt}

\makeatletter
\pgfkeys{%
  /tikz/on layer/.code={
    \pgfonlayer{#1}\begingroup
    \aftergroup\endpgfonlayer
    \aftergroup\endgroup
  },
  /tikz/node on layer/.code={
    \gdef\node@@on@layer{%
      \setbox\tikz@tempbox=\hbox\bgroup\pgfonlayer{#1}\unhbox\tikz@tempbox\endpgfonlayer\pgfsetlinewidth{\thickness}\egroup}
    \aftergroup\node@on@layer
  },
  /tikz/end node on layer/.code={
    \endpgfonlayer\endgroup\endgroup
  }
}
\def\node@on@layer{\aftergroup\node@@on@layer}
\makeatother

\makeatletter
\pgfdeclareshape{state}
{
    \savedanchor\centerpoint
    {
        \pgf@x=0pt
        \pgf@y=0pt
    }
    \anchor{center}{\centerpoint}
    \anchorborder{\centerpoint}
    \saveddimen\overallwidth
    {
        \pgf@x=3\wd\pgfnodeparttextbox
        \ifdim\pgf@x<\minimumstatewidth
            \pgf@x=\minimumstatewidth
        \fi
    }
    \savedanchor{\upperrightcorner}
    {
        \pgf@x=.5\wd\pgfnodeparttextbox
        \pgf@y=.5\ht\pgfnodeparttextbox
        \advance\pgf@y by -.5\dp\pgfnodeparttextbox
    }
    \anchor{A}
    {
        \pgf@x=-\overallwidth
        \divide\pgf@x by 4
        \pgf@y=0pt
    }
    \anchor{B}
    {
        \pgf@x=\overallwidth
        \divide\pgf@x by 4
        \pgf@y=0pt
    }
    \anchor{text}
    {
        \upperrightcorner
        \pgf@x=-\pgf@x
        \ifhflip
            \pgf@y=-\pgf@y
            \advance\pgf@y by 0.4\stateheight
        \else
            \pgf@y=-\pgf@y
            \advance\pgf@y by -0.4\stateheight
        \fi
    }
    \backgroundpath
    {
       \begin{pgfonlayer}{foreground}
        \pgfsetstrokecolor{black}
        \pgfsetlinewidth{\thickness}
        \pgfpathmoveto{\pgfpoint{-0.5*\overallwidth}{0}}
        \pgfpathlineto{\pgfpoint{0.5*\overallwidth}{0}}
        \ifhflip
            \pgfpathlineto{\pgfpoint{0}{\stateheight}}
        \else
            \pgfpathlineto{\pgfpoint{0}{-\stateheight}}
        \fi
        \pgfpathclose
        \ifblack
            \pgfsetfillcolor{black!50}
            \pgfusepath{fill,stroke}
        \else
            \pgfusepath{stroke}
        \fi
       \end{pgfonlayer}
    }
}

\pgfdeclareshape{my ground}
{
  \inheritsavedanchors[from=rectangle ee]
  \inheritanchor[from=rectangle ee]{center}
  \inheritanchor[from=rectangle ee]{north}
  \inheritanchor[from=rectangle ee]{south}
  \inheritanchor[from=rectangle ee]{east}
  \inheritanchor[from=rectangle ee]{west}
  \inheritanchor[from=rectangle ee]{north east}
  \inheritanchor[from=rectangle ee]{north west}
  \inheritanchor[from=rectangle ee]{south east}
  \inheritanchor[from=rectangle ee]{south west}
  \inheritanchor[from=rectangle ee]{input}
  \inheritanchor[from=rectangle ee]{output}
  \anchorborder{\csname pgf@anchor@my ground@input\endcsname}

  \backgroundpath{
    \pgf@process{\pgfpointadd{\southwest}{\pgfpoint{\pgfkeysvalueof{/pgf/outer xsep}}{\pgfkeysvalueof{/pgf/outer ysep}}}}
    \pgf@xa=\pgf@x \pgf@ya=\pgf@y
    \pgf@process{\pgfpointadd{\northeast}{\pgfpointscale{-1}{\pgfpoint{\pgfkeysvalueof{/pgf/outer xsep}}{\pgfkeysvalueof{/pgf/outer ysep}}}}}
    \pgf@xb=\pgf@x \pgf@yb=\pgf@y
    \pgfpathmoveto{\pgfqpoint{\pgf@xa}{\pgf@ya}}
    \pgfpathlineto{\pgfqpoint{\pgf@xa}{\pgf@yb}}
    \pgfmathsetlength\pgf@xa{.5\pgf@xa+.5\pgf@xb}
    \pgfmathsetlength\pgf@yc{.16666\pgf@yb-.16666\pgf@ya}
    \advance\pgf@ya by\pgf@yc
    \advance\pgf@yb by-\pgf@yc
    \pgfpathmoveto{\pgfqpoint{\pgf@xa}{\pgf@ya}}
    \pgfpathlineto{\pgfqpoint{\pgf@xa}{\pgf@yb}}
    \advance\pgf@ya by\pgf@yc
    \advance\pgf@yb by-\pgf@yc
    \pgfpathmoveto{\pgfqpoint{\pgf@xb}{\pgf@ya}}
    \pgfpathlineto{\pgfqpoint{\pgf@xb}{\pgf@yb}}
  }
}
\makeatother

\tikzset{inline text/.style =
  {text height=1.2ex,text depth=0.25ex,yshift=0.5mm}}
\tikzset{arrow box/.style =
  {rectangle,inline text,fill=white,draw,
    minimum height=5mm,yshift=-0.5mm,minimum width=5mm}}
\tikzset{bubble/.style =
  {inner sep=0mm,minimum width=3mm,minimum height=3mm,
    draw,shape=circle,fill=white}}
\tikzset{dot/.style =
  {inner sep=0mm,minimum width=1mm,minimum height=1mm,
    draw,shape=circle}}
\tikzset{white dot/.style = {dot,fill=white,text depth=-0.2mm}}
\tikzset{scalar/.style = {diamond,draw,inner sep=1pt}}
\tikzset{square/.style =
  {inner sep=0mm,minimum width=2mm,minimum height=2mm,
    draw,shape=rectangle}}

\tikzset{star/.style = {dot,fill=white,text depth=-0.2mm}}
\tikzset{copier/.style = {dot,fill,text depth=-0.2mm}}
\tikzset{fakecopier/.style = {square,fill,text depth=-0.2mm}}
\tikzset{discarder/.style = {my ground,draw,inner sep=0pt,
    minimum width=4.2pt,minimum height=11.2pt,anchor=input,rotate=90}}

\tikzset{xshiftu/.style = {shift = {(#1, 0)}}}
\tikzset{yshiftu/.style = {shift = {(0, #1)}}}
\tikzset{scriptstyle/.style={font=\everymath\expandafter{\the\everymath\scriptstyle}}}

\title{The information loss of a stochastic map}
\author{James Fullwood and Arthur J.~Parzygnat}

\newcommand{\Addresses}{{
  \bigskip
  \footnotesize

  A.~Parzygnat, \textsc{Institut des Hautes \'Etudes Scientifiques, 35 Route de Chartres 91440, Bures-sur-Yvette, France}\par\nopagebreak
  \textit{E-mail address}, A.~Parzygnat: \texttt{parzygnat@ihes.fr}

  \medskip

  J.~Fullwood, \textsc{School of Mathematical Sciences, Shanghai Jiao Tong University, 800 Dongchuan Road, Shanghai 200240, China}\par\nopagebreak
  \textit{E-mail address}, J.~Fullwood: \texttt{fullwood@sjtu.edu.cn}

}}

\begin{document}
\emergencystretch 2em

\maketitle
\begin{abstract}  
We provide a stochastic extension of the Baez--Fritz--Leinster characterization of the Shannon information loss associated with a measure-preserving function. This recovers the conditional entropy and a closely related information-theoretic measure that we call \emph{conditional information loss}. Although not functorial, these information measures are  semi-functorial, a concept we introduce that is definable in any Markov category. We also introduce the notion of an \emph{entropic Bayes' rule} for information measures, and we provide a characterization of conditional entropy in terms of this rule. 
\blfootnote{\emph{2020 Mathematics Subject Classification.} 
Primary 94A17; Secondary 18A05, 62F15}
\blfootnote{
\emph{Key words and phrases.} 
Bayes; conditional probability; disintegration; entropy; error correction; functor; information theory; Markov category; stochastic map; synthetic probability}
\end{abstract}

\vspace{-7mm}
\tableofcontents

\section{Introduction}

The information loss $K(f)$ associated with a measure-preserving function $(X,p)\xrightarrow{f} (Y,q)$ between finite probability spaces is given by the Shannon entropy difference 
\[
K(f):=H(p)-H(q), 
\]
where $H(p):=-\sum_{x\in X}p_{x}\log p_{x}$ is the Shannon entropy of $p$ (and similarly for $q$). 
In~\cite{BFL}, Baez, Fritz, and Leinster proved that the information loss satisfies, and is uniquely characterized up to a multiplicative factor by, the following conditions:
\begin{enumerate}
\itemsep0pt
\setcounter{enumi}{-1}
\item
\underline{\smash{Positivity}}: $K(f)\ge0$ for all $(X,p)\xrightarrow{f}(Y,q)$. This says that the information loss associated with a deterministic process is always non-negative. 
\item
\underline{\smash{Functoriality}}: $K(g\circ f)=K(g)+K(f)$ for every composable pair $(f,g)$ of measure-preserving maps. This says that the information loss of two successive processes is the sum of the information losses associated with each process. 
\item
\underline{\smash{Convex Linearity}}: $K(\lambda f\oplus (1-\lambda)g)=\lambda K(f)+(1-\lambda)K(g)$ for all $\lambda \in (0,1)$. This says that the information loss associated with tossing a (possibly unfair) coin in deciding amongst two processes is the associated weighted sum of their information losses. 
\item
\underline{\smash{Continuity}}: $K(f)$ is a continuous function of $f$. This says that the information loss does not change much under small perturbations (i.e., is robust with respect to errors).
\end{enumerate}  

As measure-preserving functions may be viewed as deterministic stochastic maps, it is natural to ask whether there exist extensions of the Baez--Fritz--Leinster (BFL) characterization of information loss to maps that are inherently random (i.e., stochastic) in nature. In particular, what information-theoretic quantity captures such an information loss in this larger category?

This question is answered in the present work. Namely, we extend the BFL characterization theorem, which is valid on \emph{deterministic} maps, to the larger category of \emph{stochastic} maps. In doing so, we also find a characterization of the conditional entropy. Although the resulting extension is not functorial on the larger category of stochastic maps, we formalize a weakening of functoriality that restricts to functoriality on deterministic maps. This weaker notion of functoriality is definable in any Markov category~\cites{ChJa18,Fr19}, and it provides a key axiom in our characterization. 

To explain how we arrive at our characterization, let us first recall the definition of stochastic maps between finite probability spaces, for which the measure-preserving functions are a special case. A \define{stochastic map} $(X,p)\xstoch{f} (Y,q)$ associates with every $x\in X$ a probability distribution $f_x$ on $Y$ such that $q_y=\sum_{x\in X}f_{yx}p_x$, where $f_{yx}$ is the distribution $f_x$ evaluated at $y\in Y$. In terms of information flow, the space $(X,p)$ may be thought of as a probability distribution on the set of inputs for a communication channel described by the stochastic matrix $f_{yx}$, while $(Y,q)$ is then thought of as the induced distribution on the set of outputs of the channel. 

Extending the information loss functor by assigning $H(p)-H(q)$ to any \emph{stochastic} map $(X,p)\xstoch{f}(Y,q)$ would indeed result in an assignment that satisfies conditions 1, 2, and 3 listed above. However, it would no longer be positive and the interpretation as an information loss would be gone. Furthermore, no additional information about the stochasticity of the map $f$ would be used in determining this assignment. In order to guarantee positivity, an additional term, depending on the stochasticity of $f$, is needed. This term is provided by the \define{conditional entropy} of $(X,p)\xstoch{f}(Y,q)$ and is given by the the non-negative real number 
\[
H(f|p):=\sum_{x\in X}p_x H(f_x),
\]
where $H(f_x)$ is the Shannon entropy of the distribution $f_x$ on $Y$.%
\footnote{In the case that $(X,p)$ and $(Y,q)$ are probability spaces associated with the alphabets of random variables $\mathcal{X}$ and $\mathcal{Y}$, then $H(f|p)$ coincides with conditional entropy $H(\mathcal{Y}|\mathcal{X})$~\cite{CoTh06}.}
If $(X,p)\xstoch{f}(Y,p)$ is in fact deterministic, i.e., if $f_x$ is a point-mass distribution for all $x\in X$, then $H(f_x)=H(f|p)=0$ for all $x\in X$. As such, $H(f|p)$ is a measure of the uncertainty (or randomness) of the outputs of $f$ averaged over the prior distribution $p$ on the set $X$ of its inputs. Indeed, $H(f|p)$ is maximized precisely when $f_x$ is the uniform distribution on $Y$ for all $x\in X$. 

Therefore, given a stochastic map $(X,p)\xstoch{f} (Y,q)$, we call
\[
K(f):=H(p)-H(q)+H(f|p)
\]
the \define{conditional information loss} of $(X,p)\xstoch{\!f}(Y,q)$.%
\footnote{The same letter $K$ is used here because it agrees with the Shannon entropy difference when $f$ is deterministic.}
As $H(f|p)=0$ whenever $f$ is deterministic, the conditional information loss restricts to the category of measure preserving functions as the information loss functor of Baez, Fritz, and Leinster, while also satisfying conditions 0,~2, and 3 (i.e., positivity, convex linearity, and continuity) on the larger category of stochastic maps. However, conditional information loss is \emph{not} functorial in general, and while this may seem like a defect at first glance, we prove that there is no extension of the information loss functor that remains functorial on the larger category of stochastic maps if the positivity axiom is to be preserved, thus retaining an interpretation as information loss. In spite of this, conditional information loss does satisfy a weakened form of functoriality, which we briefly describe now. 

A pair $(X,p)\xstoch{f}(Y,q)\xstoch{g}(Z,r)$ of composable stochastic maps is \define{a.e.\ coalescable} if and only if for every pair of elements $z\in Z$ and $x\in X$ for which $r_{z}>0$ and $p_{x}>0$, there exists a unique $y\in Y$ such that $f_{yx}>0$ and $g_{zy}>0$. Intuitively, this says that the information about the intermediate step can be recovered given knowledge about the input and output. In particular, if $f$ is deterministic, then the pair $(f,g)$ is a.e.\ colescable (for obvious reasons, since knowing $x$ alone is enough to determine the intermediate value). However, there are other many situations where a pair could be a.e.\ coalescable and the maps need not be deterministic. With this definition in place (which we also generalize to the setting of arbitrary Markov categories), we replace functoriality with the following weaker condition. 
\begin{itemize}
\item[$1^{\star}$.]
\underline{\smash{Semi-functoriality}}: $K(g\circ f)=K(g)+K(f)$ for every a.e.\ coalescable pair $(X,p)\xstoch{f}(Y,q)\xstoch{g}(Z,r)$ of stochastic maps. This says that the conditional information loss of two successive processes is the sum of the conditional information losses associated with each process \emph{provided} that the information in the intermediate step can always be recovered.
\end{itemize}

\begin{figure}
\centering
\begin{subfigure}{0.44\textwidth}
\centering
\begin{tikzpicture}[decoration=snake,scale=0.75]
\def\ra{0.56419}
\def\rb{0.56419}
\def\rc{0.56419}
\def\raa{0.281962}
\def\rba{0.322183}
\def\rca{0}
\def\rda{0.367431}
\def\rab{0}
\def\rbb{0.470071}
\def\rcb{0.163219}
\def\rdb{0.265899}
\def\rac{0.219826}
\def\rbc{0.235131}
\def\rcc{0.00695653}
\def\rdc{0.463305}
\def\sa{0.357528}
\def\sb{0.616486}
\def\sc{0.163367}
\def\sd{0.648351}
\node at (-3,-1) {$X$};
\draw[blue,thin,fill=blue,fill opacity=0.4] (-3,3) circle (\ra);
\draw[blue,thin,fill=blue,fill opacity=0.4] (-3,1.5) circle (\rb);
\draw[blue,thin,fill=blue,fill opacity=0.4] (-3,0) circle (\rc);
\node at (2.25,-1) {$X\times Y$};
\draw[blue,thin,fill=blue,fill opacity=0.4] (0,3) circle (\raa);
\draw[blue,thin,fill=blue,fill opacity=0.4] (1.5,3) circle (\rba);
\draw[blue,thin,fill=blue,fill opacity=0.4] (3,3) circle (\rca);
\draw[blue,thin,fill=blue,fill opacity=0.4] (4.5,3) circle (\rda);
\draw[blue,thin,fill=blue,fill opacity=0.4] (0,1.5) circle (\rab);
\draw[blue,thin,fill=blue,fill opacity=0.4] (1.5,1.5) circle (\rbb);
\draw[blue,thin,fill=blue,fill opacity=0.4] (3,1.5) circle (\rcb);
\draw[blue,thin,fill=blue,fill opacity=0.4] (4.5,1.5) circle (\rdb);
\draw[blue,thin,fill=blue,fill opacity=0.4] (0,0) circle (\rac);
\draw[blue,thin,fill=blue,fill opacity=0.4] (1.5,0) circle (\rbc);
\draw[blue,thin,fill=blue,fill opacity=0.4] (3,0) circle (\rcc);
\draw[blue,thin,fill=blue,fill opacity=0.4] (4.5,0) circle (\rdc);
\draw[-{>[scale=2.5,
          length=2,
          width=3]},decorate,thick] (-2.0,1.5) -- node[below,xshift=0.1cm]{$\bloom_{f}$} (-0.5,1.5);
\node at (0,3.85) {};
\node at (0,-2.5) {};
\end{tikzpicture}
\caption{The bloom of $X\xstoch{f}Y$}
\label{fig:bloom}
\end{subfigure}
\quad
\begin{subfigure}{0.52\textwidth}
\centering
\begin{tikzpicture}[decoration=snake,scale=0.75]
\def\ra{0.217079}
\def\rb{0.411546}
\def\rc{0.319086}
\def\raa{0.281962*1.77245*\ra}
\def\rba{0.322183*1.77245*\ra}
\def\rca{0*1.77245*\ra}
\def\rda{0.367431*1.77245*\ra}
\def\rab{0*1.77245*\rb}
\def\rbb{0.470071*1.77245*\rb}
\def\rcb{0.163219*1.77245*\rb}
\def\rdb{0.265899*1.77245*\rb}
\def\rac{0.219826*1.77245*\rc}
\def\rbc{0.235131*1.77245*\rc}
\def\rcc{0.00695653*1.77245*\rc}
\def\rdc{0.463305*1.77245*\rc}
\def\sa{0.165006}
\def\sb{0.388106}
\def\sc{0.119125}
\def\sd{0.35534}
\node at (-3,-1) {$(X,p)$};
\draw[blue,thin,fill=blue,fill opacity=0.4] (-3,3) circle (\ra);
\draw[blue,thin,fill=blue,fill opacity=0.4] (-3,1.5) circle (\rb);
\draw[blue,thin,fill=blue,fill opacity=0.4] (-3,0) circle (\rc);
\draw[blue,thin,fill=blue,fill opacity=0.4] (0,3) circle (\raa);
\draw[blue,thin,fill=blue,fill opacity=0.4] (1.5,3) circle (\rba);
\draw[blue,thin,fill=blue,fill opacity=0.4] (3,3) circle (\rca);
\draw[blue,thin,fill=blue,fill opacity=0.4] (4.5,3) circle (\rda);
\draw[blue,thin,fill=blue,fill opacity=0.4] (0,1.5) circle (\rab);
\draw[blue,thin,fill=blue,fill opacity=0.4] (1.5,1.5) circle (\rbb);
\draw[blue,thin,fill=blue,fill opacity=0.4] (3,1.5) circle (\rcb);
\draw[blue,thin,fill=blue,fill opacity=0.4] (4.5,1.5) circle (\rdb);
\draw[blue,thin,fill=blue,fill opacity=0.4] (0,0) circle (\rac);
\draw[blue,thin,fill=blue,fill opacity=0.4] (1.5,0) circle (\rbc);
\draw[blue,thin,fill=blue,fill opacity=0.4] (3,0) circle (\rcc);
\draw[blue,thin,fill=blue,fill opacity=0.4] (4.5,0) circle (\rdc);
\node at (-1.0,-3) {$(Y,q)$};
\draw[blue,thin,fill=blue,fill opacity=0.4] (0,-3) circle (\sa);
\draw[blue,thin,fill=blue,fill opacity=0.4] (1.5,-3) circle (\sb);
\draw[blue,thin,fill=blue,fill opacity=0.4] (3,-3) circle (\sc);
\draw[blue,thin,fill=blue,fill opacity=0.4] (4.5,-3) circle (\sd);
\draw[-{>[scale=2.5,
          length=2,
          width=3]},thick] (2.25,-1.0) -- node[right]{$\pi_{Y}$} (2.25,-2.0);
\draw[-{>[scale=2.5,
          length=2,
          width=3]},decorate,thick] (-2.0,1.5) -- node[below,xshift=0.1cm]{$\bloom_{f}$} (-0.5,1.5);
\draw[-{>[scale=2.5,
          length=2,
          width=3]},decorate,thick] (-2.0,-0.5) -- node[below,xshift=-0.1cm]{$f$} (-0.5,-2.0);
\end{tikzpicture}
\caption{The bloom-shriek factorization of $(X,p)\xstoch{f}(Y,q)$}
\label{fig:bloomshriek}
\end{subfigure}
\caption{A visualization of bloom and the bloom-shriek factorization via water droplets as inspired by Gromov~\cite{Gr14}. The bloom of $f$ splits each water droplet of volume 1 (an element of $X$) into several water droplets whose total volume equates to 1. If $X$ has a probability $p$ on it, then the initial volume of that water droplet is scaled by this probability. The stochastic map therefore splits the water droplet using this scale.}
\end{figure}
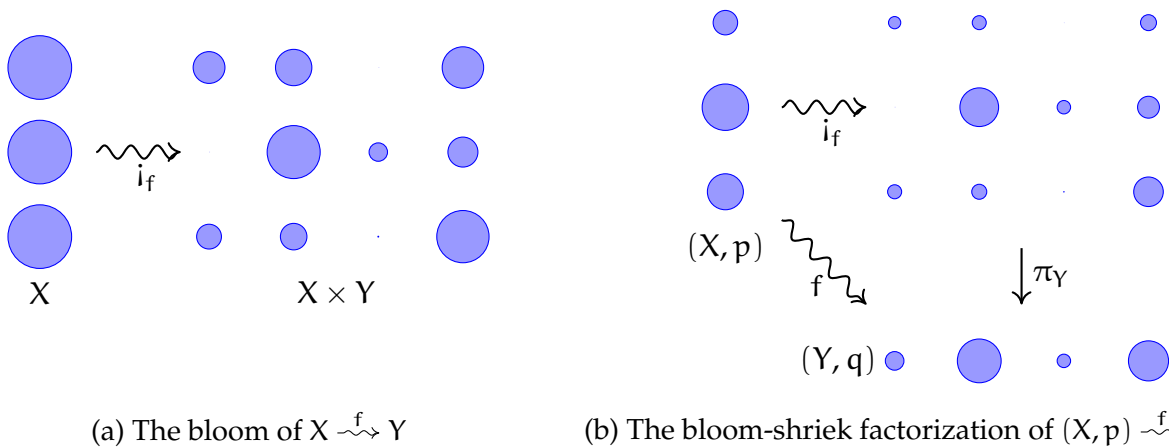

Replacing functoriality with semi-functoriality is not enough to characterize the conditional information loss. However, it comes quite close, as only one more axiom is needed. Assuming positivity, semi-functoriality, convex linearity, and continuity, there are several equivalent axioms that may be stipulated to characterize the conditional information loss. To explain the first option, we introduce a convenient factorization of every stochastic map $(X,p)\xstoch{f}(Y,q)$. The \define{bloom-shriek factorization} of $f$ is given by the decomposition $f=\pi_{Y}\circ\bloom_{f}$, where $\bloom_{f}:X\xstoch{}X\times Y$ is the \define{bloom} of $f$ whose value at $x'$ is the probability measure on $X\times Y$ given by sending $(x,y)$ to $\delta_{x'x}f_{yx'}$, where $\delta_{x'x}$ is the Kronecker delta. In other words, $\bloom_{f}$ records each of the probability measures $f_{x}$ on a copy of $Y$ indexed by $x\in X$. A visualization of the bloom of $f$ is given in Figure~\ref{fig:bloom}. When one is given the additional data of probability measures $p$ and $q$ on $X$ and $Y$, respectively, then Figure~\ref{fig:bloomshriek} illustrates the bloom-shriek factorization of $f$. From this point of view, $\bloom_{f}$ keeps track of the information encoded in \emph{both} $p$ and $f$, while the projection map $\pi_{Y}$ forgets, or loses, \emph{some} of this information. 

With this in mind, our final axiom to characterize the conditional information loss is 
\begin{itemize}
\item[4 (a).]
\underline{Reduction}: $K(f)=K(\pi_{Y})$, where $f=\pi_{Y}\circ\bloom_{f}$ is the bloom-shriek factorization of $f$. This says that the conditional information loss of $f$ equals the information loss of the projection using the associated joint distribution on $X\times Y$. 
\end{itemize}
Note that this axiom describes how $K$ is determined by its action on an associated class of deterministic morphisms. 
These slightly modified axioms, namely, semi-functoriality, convex linearity, continuity, and reduction, characterize the conditional information loss and therefore extend Baez, Fritz, and Leinster's characterization of information loss. A much simpler axiom that may be invoked in place of the reduction axiom which also characterizes conditional information loss is the following. 
\begin{itemize}
\item[4 (b).]
\underline{\smash{Blooming}}: $K(\bloom_p)=0$, where $\bloom_p$ is the unique map $(\bullet,1)\xstoch{}(X,p)$ from a one point probability space to $(X,p)$. This says that if a process begins with no prior information, then there is no information to be lost in the process. 
\end{itemize}

The conditional entropy itself can be extracted from the conditional information loss by a process known as Bayesian inversion, which we now briefly recall. Given a stochastic map $(X,p)\xstoch{f}(Y,q)$, there exists a stochastic map $(Y,q)\xstoch{\ov f}(Y,q)$ such that $f_{yx}p_{x}=\ov{f}_{xy}q_{y}$ for all $x\in X$ and $y\in Y$ (the stochastic map $\ov f$ is the almost everywhere unique conditional probability so that Bayes' rule holds). Such a map is called a \define{Bayesian inverse} of $f$. The Bayesian inverse can be visualized using the bloom-shriek factorization because it itself has a bloom-shriek factorization $\ov f=\pi_{X}\circ \bloom_{\ov f}$. 
This is obtained by finding the stochastic maps in the opposite direction of the arrows so that they reproduce the appropriate volumes of the water droplets. 

Given this perspective on Bayesian inversion, we prove that the \emph{conditional entropy} of $(X,p)\xstoch{f}(Y,q)$ equals the \emph{conditional information loss} of its Bayesian inverse $(Y,q)\xstoch{\ov f}(X,p)$. And since the conditional information loss of $\ov f$ is just the information loss of $\pi_{X}$, this indicates how the conditional entropy and conditional information losses are the ordinary information losses associated with the two projections $\pi_{X}$ and $\pi_{Y}$ in Figure~\ref{fig:bloomshriek}. 
This duality also provides an interesting perspective on conditional entropy and its characterization. 
Indeed, using Bayesian inversion, we also characterize the \emph{conditional entropy} as the unique assignment $F$ sending measure-preserving stochastic maps between finite probability spaces to real numbers satisfying conditions 0, $1^{\star}$, 2, and 3 above, but with a new axiom 
that reads as follows. 
\begin{itemize}
\item[4 (c).]
\underline{\smash{Entropic Bayes' Rule}}: $F(f)+F(\bloom_{p})=F(\ov{f})+F(\bloom_{q})$ for all $(X,p)\xstoch{f}(Y,q)$. This is an information theoretic analogue of Bayes' rule, which reads $f_{yx} p_{x}=\ov{f}_{xy} q_{y}$ for all $x\in X$ and $y\in Y$, or in more traditional probabilistic notation $\mathbb{P}(y|x)\mathbb{P}(x)=\mathbb{P}(x|y)\mathbb{P}(y).$
\end{itemize}
In other words, we obtain a \emph{Bayesian} characterization of the conditional entropy.  
This provides an entropic and information-theoretic description of Bayes' rule from the Markov category perspective, in a way that we interpret as answering an open question of Fritz~\cite{Fr20}. 

\section{Categories of stochastic maps}
\label{sec:FinPS}

In the first few sections, we define all the concepts involved in proving that the conditional information loss satisfies the properties that we will later prove characterize it. This section introduces the domain category and its convex structure. 

\bd
Let $X$ and $Y$ be finite sets.  
A \define{stochastic map} $f:X\xstoch{} Y$  associates a probability measure 
$f_{x}$ to every $x\in X$. If $f:X\xstoch{} Y$ is such that $f_x$ is a point-mass distribution for every $x\in X$, then $f$ is said be to \define{deterministic}.
\ed

\bnot
Given a stochastic map $f:X\xstoch{} Y$ (also written as $X\xstoch{f} Y$), the value $f_x(y)\in [0,1]$ will be denoted by $f_{yx}$.
As there exists a canonical bijection between deterministic maps of the form $X\xstoch{} Y$ and functions $X\to Y$, deterministic maps from $X$ to $Y$ will be denoted by the functional notation $X\to Y$.
\enot

\bd
A stochastic map of the form $\bullet\xstoch{} X$ from a single element set to a finite set $X$ is a single probability measure on $X$. Its unique value at $x$ will be denoted by $p_x$ for all $x\in X$. The set
$
N_p:=\{x\in X\hspace{1mm}|\hspace{1mm} p_x=0\}
$
will be referred to as the \define{nullspace} of $p$.
\ed

\bd
Let $\mathbf{FinStoch}$ be the category of stochastic maps between finite sets. Given a finite set $X$, the identity map of $X$ in $\mathbf{FinStoch}$  corresponds to the identity \emph{function} $\text{id}_X:X\to X$. Second, given stochastic maps $f:X\xstoch{} Y$ and $g:Y\xstoch{} Z$, the composite $g\circ f:X\xstoch{} Z$ is given by the Chapmann--Kolmogorov equation
$
(g\circ f)_{zx}:=\sum_{y\in Y}g_{zy}f_{yx}.
$
\ed

\bd
\label{DSM77}
Let $X$ be a finite set. The \define{copy} of $X$ is the diagonal embedding $\Delta_X:X\to X\times X$, and the \define{discard} of $X$ is the unique map from $X$ to the terminal object $\bullet$ in $\mathbf{FinStoch}$, which will be denoted by $\shriek_X:X\to \bullet$. 
If $Y$ is another finite set, the \define{swap map} is the map $\gamma:X\times Y\to Y\times X$ given by $(x,y)\mapsto (y,x)$.
Given morphisms $f:X\xstoch{} X'$ and $g:Y\xstoch{} Y'$ in $\FinStoch$,  the \define{product} of $f$ and $g$ is the stochastic map $f\times g:X\times Y\xstoch{} X'\times Y'$ given by
$
(f\times g)_{(x',y')(x,y)}:=f_{x'x}g_{y'y}.
$
\ed

The product of stochastic maps endows $\FinStoch$ with the structure of a monoidal category. Together with the copy, discard, and swap maps, $\FinStoch$ is a Markov category~\cites{ChJa18,Fr19}.
 
\bd
Let $\FinPS$ (this stands for ``\textbf{fin}ite \textbf{p}robabilities and \textbf{s}tochastic maps'') be the co-slice category $\bullet\downarrow \mathbf{FinStoch}$, i.e., the category whose objects are pairs $(X,p)$ consisting of a finite set $X$ equipped with a probability measure $p$, and a morphism from $(X,p)$ to $(Y,q)$ is a stochastic map $X\xstoch{f}Y$ such that
$\ds q_{y}=\sum_{x\in X}f_{yx}p_{x}$ for all $y\in Y$.
The subcategory of deterministic maps in $\FinPS$ will then be denoted by $\FinPD$ (which stands for ``\textbf{fin}ite \textbf{p}robabilities and \textbf{d}eterministic maps'').%
\footnote{The category $\FinPD$ was called $\FinProb$ in~\cite{BFL}.}
A pair $(f,g)$ of morphisms in $\FinPS$ is said to be a \define{composable pair} iff $g\circ f$ exists.
\ed

\br
Though it is often the case that we will denote a morphism $(X,p)\xstoch{f} (Y,q)$ in $\FinPS$ simply by $f$, such notation is potentially ambiguous, as the morphism $(X,p')\xstoch{f} (Y,q')$ is distinct from the morphsim $(X,p)\xstoch{f} (Y,q)$ whenever $p\neq p'$. As such, we will only employ the shorthand of denoting a morphism in $\FinPS$ by its underlying stochastic map whenever the source and target of the morphism are clear from the context.
\er

\blem
The object $(\bullet,1)$ given by a single element set equipped with the unique probability measure is a zero object (i.e.\ terminal and initial) in $\FinPS$.
\elem

\bd
Given an object $(X,p)$ in $\FinPS$, the \define{shriek} and \define{bloom} of $p$ are the unique maps to and from $(\bullet,1)$ respectively, which will be denoted $\shriek_p:(X,p)\to (\bullet,1)$ and $\bloom_{p}:(\bullet,1)\xstoch{} (X,p)$ (the former is deterministic, while the latter is stochastic). The underlying stochastic maps associated with $\shriek_p$ and $\bloom_p$ are $\shriek_X:X\to \bullet$ and $p:\bullet\xstoch{} X$, respectively.
\ed

\bx
\label{ex:zeroobject}
Since $(\bullet,1)$ is a zero object, given any two objects $(X,p)$ and $(Y,q)$, there exists at least one morphism $(Y,q)\xstoch{}(X,p)$, namely the composite $(Y,q)\xrightarrow{\shriek_{q}}(\bullet,1)\xstoch{\bloom_{p}}(X,p)$. 
\ex

\bd
\label{defn:jointdistribution}
Let $(X,p)\xstoch{f} (Y,q)$ be a morphism in $\FinPS$. The \define{joint distribution} associated with $f$ is the probability measure $\vartheta(f):\bullet\xstoch{} X\times Y$ given by $\vartheta(f)_{(x,y)}=f_{yx}p_x$. 
\ed

It is possible to take convex combinations of both objects and morphisms in $\FinPS$, and such assignments will play a role in our characterization of conditional entropy.

\bd
\label{defn:convexcombos}
Let $p:\bullet\xstoch{} X$ be a probability measure and let $\{(Y_{x},q^{x})\}_{x\in X}$ be a collection of objects in $\mathbf{FinPS}$ indexed by $X$. The $p$-\define{weighted convex sum} $\bigoplus_{x\in X}p_{x}(Y_{x},q^{x})$ is defined to be the set%
\footnote{This set is well-defined up to canonical isomorphism by the universal property of the disjoint union. We will often ignore these canonical isomorphisms.}
$\coprod_{x\in X}Y_{x}$ equipped with the probability measure $\bigoplus_{x\in X}p_{x}q^{x}$ given by
\[
\left(\bigoplus_{x\in X} p_xq^x\right)_z:=
p_x q^x_z \quad \hspace{0.15cm} \text{if } z\in Y_x.
\]
In addition, if $(Y_x,q^x)\xstoch{Q^x} (Y'_x,q'^x)$ is a collection of morphisms in $\mathbf{FinPS}$ indexed by $X$, the $p$-\define{weighted convex sum} $\bigoplus_{x\in X} p_xQ^x:\left(\coprod_{x\in X}Y_x,\bigoplus_{x\in X} p_xq^x\right)\xstoch{} \left(\coprod_{x\in X}Y'_x,\bigoplus_{x\in X} p_xq'^x\right)$ is given by
\[
\left(\bigoplus_{x\in X} p_{x} Q^x\right)_{z'z}:=
\begin{cases}
Q^x_{z'z} & \mbox{if $(z',z)\in Y'_x\times Y_x$} \\
0 & \mbox{otherwise.} \\
\end{cases}
\]
\ed

\section{The Baez--Fritz--Leinster characterization of information loss}
\label{sec:BFL}

In \cite{BFL}, Baez, Fritz, and Leinster (BFL) characterized the Shannon entropy difference associated with measure-preserving functions between finite probability spaces as the only non-vanishing, continuous, convex linear functor from $\FinPD$ to the non-negative reals (up to a multiplicative constant).  It is then natural to ask whether there exist either extensions or analogues of their result by including non-deterministic morphisms from the larger category $\FinPS$. Before delving deeper into such inquiry, we first recall in detail the characterization theorem of BFL. 

\bd
Let $\B\R$ be the convex category consisting of a single object and whose set of morphisms is $\R$. The composition in $\B\R$ is given by addition. Convex combinations of morphisms are given by ordinary convex combinations of numbers. The subcategory of non-negative reals will be denoted $\B\R_{\ge0}$. 
\ed 

\begin{convention}
\label{conv:functions}
In the rest of the paper, we will not necessarily assume that assignments from one category to another are functors. Nevertheless, we do assume they form (class) functions (see~\cite[Section~I.7]{Ma98} for more details). Furthermore, we assume that they respect or reflect source and targets in the following sense. If $\mC$ and $\mD$ are two categories, all functions $F:\mC\to\mD$ are either \define{covariant} or \define{contravariant} in the sense that for any morphism $a\xrightarrow{\gamma}b$ in $\mC$, $F(\gamma)$ is a morphism from $F(a)$ to $F(b)$ or from $F(b)$ to $F(a)$, respectively. These are the only types of functions between categories we will consider in this work. As such, we therefore abuse terminology and use the term \emph{functions} for such assignments throughout. If $M$ is a \emph{commutative} monoid and $\B M$ denotes its one object category, then every covariant function $\mC\to\B M$ is also contravariant and vice-versa. 
\end{convention}

We now define a notion of continuity for functions of the form $F:\FinPS\to \B\R$. 

\bd
\label{defn:sequenceinFinPS}
A sequence of morphisms 
$
(X_{n},p_{n})\xstoch{f_{n}}(Y_{n},q_{n})
$
in $\FinPS$ \define{converges}
to a morphism $(X,p)\xstoch{f}(Y,q)$ if and only if the following two conditions hold.
\begin{enumerate}[(a)]
\itemsep0pt
\item
There exists an $N\in\N$ for which $X_{n}=X$ and $Y_{n}=Y$ for all $n\ge N$.
\item
The following limits%
\footnote{Note that these limits necessarily imply $\ds\lim_{n\rightarrow\infty}q_{n}=q$.}
hold:
$\ds\lim_{n\rightarrow\infty}p_{n}=p$ 
and
$\ds\lim_{n\to\infty}f_{n}=f$.
\end{enumerate}
A function $F:\FinPS\to\B\R$ is \define{continuous} if and only if
$\ds\lim_{n\rightarrow\infty}F(f_{n})=F(f)$
whenever $\{f_{n}\}$ is a sequence in $\FinPS$ converging to $f$. 
\ed

\br
In the subcategory $\FinPD$, since the topology of the collection of \emph{functions} from a finite set $X$ to another finite set $Y$ is \emph{discrete}, one can equivalently assume that a sequence $f_{n}$ as in Definition~\ref{defn:sequenceinFinPS}, but this time with all $f_{n}$ deterministic, converges to $(X,p)\xrightarrow{f}(Y,q)$ if and only if the following two conditions hold.
\begin{enumerate}[(a)]
\itemsep0pt
\item
There exists an $N\in\N$ for which $X_{n}=X, Y_{n}=Y,$ $f_{n}=f$ for all $n\ge N$.
\item
For $n\ge N$, one has
$\ds\lim_{n\rightarrow\infty}p_{n}=p$.
\end{enumerate}
In this way, our definition of convergence agrees with the definition of convergence of BFL on the subcategory $\FinPD$~\cite{BFL}.
\er

\bd
A function $F:\FinPS\to \B\R$ is said to be \define{convex linear} if and only if for all objects $(X,p)$ in $\FinPS$, 
\[
F\left(\bigoplus_{x\in X}p_{x}Q^{x}\right)=\sum_{x\in X}p_{x}F(Q^{x})
\]
for all collections $\big\{(Y_{x},q^{x})\xstoch{Q^{x}}(Y'_{x},q'^{x})\big\}_{x\in X}$ in $\FinPS$. 
\ed

\bd
A function $F:\FinPS\to \B\R$ is said to be \define{functorial} if and only if it is in fact a functor, i.e., if and only if $F(g\circ f)=F(f)+F(g)$ for every composable pair $(f,g)$ in $\FinPS$. 
\ed

\bd
\label{defn:Shannon}
Let $p:\bullet\xstoch{} X$ be a probability measure. The \define{Shannon entropy} of $p$ is given by
\[
H(p):=-\sum_{x\in X}p_{x}\log(p_{x}).
\]
\ed

\begin{convention}
When considering any entropic quantity, we will always adhere to the convention that $0\log(0)=0$.
\end{convention}

\bd
\label{defn:infolossfunctor}
Given a map $(X,p)\xrightarrow{f} (Y,p)$ in $\FinPD$, the Shannon entropy difference $H(p)-H(q)$ will be referred to as the \define{information loss} of $f$. Information loss defines a functor $K:\FinPD\to \B\R$, henceforth referred to as the \define{information loss functor} on $\FinPD$.
\ed

\bt[Baez--Fritz--Leinster~\cite{BFL}]
\label{TBFL}
Suppose $F:\FinPD\to\B\R_{\ge0}$ is a function which satisfies the following conditions.
\begin{enumerate} 
\itemsep0pt
\item\label{X1}
$F$ is functorial.
\item\label{X2}
$F$ is convex linear.
\item\label{X3}
$F$ is continuous.
\end{enumerate}
Then $F$ is a non-negative multiple of information loss. Conversely, the information loss functor is non-negative and satisfies conditions \ref{X1}-\ref{X3}.
\et

In light of Theorem~\ref{TBFL}, it is natural to question whether or not there exists a functor $K:\FinPS\to \B\R_{\geq 0}$ that restricts to $\FinPD$ as the information loss functor. It turns out that no such non-vanishing functor exists, as we prove in the following proposition.

\bn
\label{prop:functorsvanishonFinPS}
If $F:\FinPS\to\B\R_{\ge0}$ is a functor, then $F(f)=0$ for all morphisms $f$ in $\FinPS$.
\en
\bprf
Let $(X,p)\xstoch{f} (Y,q)$ be a morphism in $\FinPS$. Since $F$ is a functor, 
\[
F(\bloom_q)=F(f\circ \bloom_p)=F(f)+F(\bloom_p) \implies 0\le F(f)=F(\bloom_q)-F(\bloom_p).
\]
Let $(Y,q)\xstoch{g}(X,p)$ be \emph{any} morphism in $\FinPS$ (which necessarily exists by Example~\ref{ex:zeroobject}, for instance). Then a similar calculation yields
\[
0\le F(g)=F(\bloom_p)-F(\bloom_q)=-F(f).
\]
Hence, $F(f)=0$.
\eprf

\section{Extending the information loss functor}
\label{SCIL}

Proposition~\ref{prop:functorsvanishonFinPS} shows it is not possible to extend the information loss functor to a functor on $\FinPS$. Nevertheless, in this section, we define a non-vanishing \emph{function} $K:\FinPS\to \B\R_{\geq 0}$ that restricts to the information loss \emph{functor} on $\FinPD$, which we refer to as \emph{conditional information loss}. While $K$ is not functorial, we show that it satisfies many important properties such as continuity, convex linearity, and invariance with respect to compositions with isomorphisms. Furthermore, in Section~\ref{sec:functorialitycondent} we show $K$ is functorial on a restricted class of composable pairs of morphisms (cf.\ Definition~\ref{coalescable}), which are definable in any Markov category. At the end of this section we characterize conditional information loss as the unique extension of the information loss functor satisfying the reduction axiom 4~(a) as stated in the introduction. In Section~\ref{CT771}, we prove an intrinsic characterization theorem for $K$ without reference to the deterministic subcategory $\FinPD$ inside $\FinPS$. Appendix~\ref{app:cccil} provides an interpretation of the vanishing of conditional information loss in terms of correctable codes.

\bd
The \define{conditional information loss} of a morphism $(X,p)\xstoch{f}(Y,q)$ in $\FinPS$ is the real number given by
\[
K(f):=H(p)-H(q)+H(f|p), 
\]
where
\[
H(f|p):=\sum_{x\in X}p_{x}H(f_{x})
\]
is  the \define{conditional entropy} of $(X,p)\xstoch{f}(Y,q)$. 
\ed

\bn
\label{prop:conditionalentropydifference}
The function $K:\FinPS\to\B\R$, uniquely determined on morphisms by sending $(X,p)\xstoch{f}(Y,q)$ to $K(f)$, 
satisfies the following conditions.
\begin{enumerate}[i)]
\itemsep0pt
\item
\label{item:Kpos}
$K(f)\ge0$.
\item
\label{item:KonFinPD}
$K$ restricted to $\FinPD$ agrees with the information loss functor (cf.\ Definition~\ref{defn:infolossfunctor}).
\item
\label{item:Kaffine}
$K$ is convex linear.
\item
\label{item:Kcont}
$K$ is continuous.
\item
\label{item:Kinfoloss}
Given $(X,p)\xstoch{f}(Y,q)$, then $K(f)=K(\pi_{Y})$, where $(X\times Y,\vartheta(f))\xrightarrow{\pi_{Y}}(Y,q)$ is the projection and $\vartheta(f)$ is the joint distribution (cf.\ Definition~\ref{defn:jointdistribution}).  
\end{enumerate}
\en

\blem
\label{lem:usefulexpressionforK}
Let $(X,p)\xstoch{f} (Y,q)$ be a morphism in $\FinPS$. Then
\[
K(f)=-\sum_{x\in X\setminus N_{p}}\sum_{y\in Y\setminus N_{f_{x}}}f_{yx}p_{x}\log\left(\frac{f_{yx}p_{x}}{q_{y}}\right).
\]
\elem

\bprf
[Proof of Lemma~\ref{lem:usefulexpressionforK}]
Applying $K$ to $f$ yields
\[
\begin{split}
K(f)&=-\sum_{x,y}f_{yx}p_{x}\log(f_{yx})-\sum_{x}p_{x}\log(p_{x})+\sum_{y}q_{y}\log(q_{y})\\
&=-\sum_{x,y}f_{yx}p_{x}\log(f_{yx})-\sum_{x}\left(\sum_{y}f_{yx}\right)p_{x}\log(p_{x})+\sum_{y}\left(\sum_{x}f_{yx}p_{x}\right)\log(q_{y})\\
&=-\sum_{x\in X\setminus N_{p}}\sum_{y\in Y\setminus N_{f_{x}}}f_{yx}p_{x}\log\left(\frac{f_{yx}p_{x}}{q_{y}}\right)
.
\qedhere
\end{split}
\]
\eprf

\bprf
[Proof of Proposition~\ref{prop:conditionalentropydifference}]
{\color{white}{you found me!}}
\begin{enumerate}[i)] 
\itemsep0pt
\item
The non-negativity of $K$ follows from Lemma~\ref{lem:usefulexpressionforK} and the equality $q_{y}=\sum_{x'\in X}f_{yx'}p_{x'}\ge f_{yx}p_{x}$.
\item
This follows from the fact that $H(f|p)=0$ for all deterministic $f$. 
\item
Let $p:\bullet\xstoch{}X$ be a probability measure, and let $(Y_x,q^x)\xstoch{Q^x} (Y'_x,q'^x)$ be a collection of morphisms in $\FinPS$ indexed by $X$. Then the $p$-weighted convex sum $\bigoplus_{x\in X}p_xQ^x$ is a morphism in $\FinPS$ of the form $(Z,r)\xstoch{h}(Z',r')$, where $Z:=\coprod_{x\in X}Y_x$, $Z':=\coprod_{x\in X}Y'_x$, $h:=\bigoplus_{x\in X}p_xQ^x$, $r:=\bigoplus_{x\in X}p_xq^x$, and $r':=\bigoplus_{x\in X}p_xq'^x$. Then
\[
\begin{split}
K(h)&=\sum_{z'}r'_{z'}\log(r'_{z'})-\sum_{z}r_{z}\log(r_{z})-\sum_{z,z'}r_zh_{z'z}\log(h_{z'z}) \\
&=\sum_{x}p_{x}\left[\sum_{y'_{x}}q'^{x}_{y'_{x}}\log(p_{x}q'^{x}_{y'_{x}})-\sum_{y_x}q^{x}_{y_{x}}\log(p_{x}q^{x}_{y_{x}})+\sum_{y_x,y'_{x}}q^{x}_{y_x}Q^{x}_{y'_{x}y_{x}}\log(Q^{x}_{y'_{x}y_{x}})\right] \\
&=\sum_{x}p_x K\left(Q^x\right),
\end{split}
\]
which shows that $K$ is convex linear.
\item
Let $\big(X^{(n)},p^{(n)}\big)\xstoch{f^{(n)}}\big(Y^{(n)},q^{(n)}\big)$ 
be a sequence (indexed by $n\in\N$) of probability-preserving stochastic maps such that $X^{(n)}=X$ and $Y^{(n)}=Y$ for large enough $n$, and where $\ds\lim_{n\to\infty}f^{(n)}=f,\lim_{n\to\infty}p^{(n)}=p,$ and $\ds\lim_{n\to\infty}q^{(n)}=q$. Then
\[
\lim_{n\to\infty}K\big(f^{(n)}\big)=-\lim_{n\to\infty}\sum_{x,y}f^{(n)}_{yx}p^{(n)}_{x}\log\left(\frac{f^{(n)}_{yx}p^{(n)}_{x}}{\sum_{x'}f^{(n)}_{yx'}p^{(n)}_{x'}}\right)
=K(f),
\]
where the last equality follows from the fact that the limit and sum (which is finite) can be interchanged and all expressions are continuous on $[0,1]$.
\item
This follows from 
\[
\begin{split}
H\big(\vartheta(f)\big)&=-\sum_{\substack{x\in X\\y\in Y}}f_{yx}p_{x}\log(f_{yx}p_{x})=-\sum_{\substack{x\in X\\y\in Y}}f_{yx}p_{x}\log(f_{yx})-\underbrace{\sum_{y\in Y}f_{yx}}_{=1}\sum_{x\in X}p_{x}\log(p_{x})\\
&=H(f|p)+H(p)
\end{split}
\]
and the fact that $K(\pi_{Y})=H\big(\vartheta(f)\big)-H(q)$. 
\qedhere 
\end{enumerate}
\eprf

\br
Since conditional entropy vanishes for deterministic morphisms, conditional information loss restricts to $\FinPD$ as the information loss functor. It is important to note that if the term $H(f|p)$ was not included in the expression for $K(f)$, then the inequality $K(f)\ge0$ would fail in general.
When $f$ is deterministic, Baez, Fritz, and Leinster proved $H(p)-H(q)\ge0$. However, when $f$ is stochastic, the inequality $H(p)-H(q)\ge0$ does not hold in general. This has to do with the fact that stochastic maps may increase entropy, whereas deterministic maps always decrease it (in the present classical setting).%
\footnote{This no longer holds for quantum systems~\cite{PaEntropy}.}
 As such, the term $H(f|p)$ is needed to retain non-negativity as one attempts to extend BFL's functor $K$ on $\FinPD$ to a function on $\FinPS$. 
\er

Item \ref{item:Kinfoloss} of Proposition~\ref{prop:conditionalentropydifference} says that the conditional information loss of a map $(X,p)\xstoch{f} (Y,q)$ in $\FinPS$ is the information loss of the deterministic map $(X\times Y,\vartheta(f))\xrightarrow{\pi_{Y}}(Y,q)$ in $\FinPD$, so that conditional information loss of a morphism in $\FinPS$ may always be reduced to the information loss of a deterministic map in $\FinPD$ naturally associated with it having the same target. This motivates the following definition. 

\bd\label{redux89}
A function $F:\FinPS\to \B\R$ is \define{reductive} if and only if $F(f)=F(\pi_Y)$ for every morphism $(X,p)\xstoch{f} (Y,q)$ in $\FinPS$ (cf.\ Proposition~\ref{prop:conditionalentropydifference} item~\ref{item:Kinfoloss} for notation).
\ed

\bn[Reductive characterization of conditional information loss]
\label{prop:prelimcharacterization}
 Let $F:\FinPS\to\B\R_{\geq 0}$ be a function satisfying the following conditions.  
 \begin{enumerate}[i)]
 \itemsep0pt
 \item\label{Z1}
 $F$ restricted to $\FinPD$ is functorial, convex linear, and continuous.
 \item\label{Z2}
 $F$ is reductive. 
 \end{enumerate}
 Then $F$ is a non-negative multiple of conditional information loss. Conversely, conditional information loss satisfies conditions \ref{Z1} and \ref{Z2}.
\en

\bprf
This follows immediately from Theorem~\ref{TBFL} and item~\ref{item:Kinfoloss} of Proposition~\ref{prop:conditionalentropydifference}. 
\eprf

In what follows, we will characterize conditional information loss without any explicit reference to the subcatgeory $\FinPD$ or the information loss functor of Baez, Fritz, and Leinster. To do this, we first need to develop some machinery.

\section{Coalescable morphisms and semi-functoriality}
\label{sec:functorialitycondent}
While conditional information loss is not functorial on $\FinPS$, we know it acts functorially on deterministic maps. As such, it is natural to ask for which pairs of composable stochastic maps does the conditional information loss act functorially. In this section, we answer this question, and then we use our result to define a property of functions $\FinPS\to \B\R$ that is a weakening of functoriality, and which we refer to as \emph{semi-functoriality}. Our definitions 
are valid in any Markov category (cf.\ Appendix~\ref{app:markovcat}). 

\bd
A deterministic map $Z\times X\xrightarrow{h}Y$ is said to be a \define{mediator} for the composable pair $(X,p)\xstoch{f} (Y,q)\xstoch{g} (Z,r)$ in $\FinPS$ if and only if 
\be
\label{MFE17}
(g\circ f)_{zx}=g_{zh(z,x)}f_{h(z,x)x} \quad \quad \text{for all} \quad (z,x)\in Z\times (X\setminus N_p).
\ee 
If in fact equation \eqref{MFE17} holds for all $(z,x) \in Z\times X$, then $h$ is said to be a \define{strong mediator} for the composable pair $X\xstoch{f} Y\xstoch{g} Z$ in $\FinStoch$.
\ed

\br
Mediators do not exist for general composable pairs, as one can see by considering any composable pair $(\bullet,1)\xstoch{p} (X,p)\xstoch {f} (Y,q)$ such that $H(\vartheta(f))\neq H(q)$ (cf.\ Definitions~\ref{defn:jointdistribution} and~\ref{defn:Shannon}).
\er

\bn
\label{prop:mediatorequivalence}
Let $(X,p)\xstoch{f} (Y,q)\xstoch{g} (Z,r)$ be a composable pair of morphisms in $\FinPS$. 
Then the following statements are equivalent.
\begin{enumerate}[(a)]
\itemsep0pt
\item
\label{item:atmostone}
For every $x\in X\setminus N_{p}$ and $z\in Z$, there exists \emph{at most} one $y\in Y$ such that $g_{zy}f_{yx}\ne0$. 
\item
\label{item:admitsmediator}
The pair $(f,g)$ admits a mediator $Z\times X\xrightarrow{h}Y$.
\item
\label{item:mediatorstringexplicit}
There exists a function $Z\times X\xrightarrow{h}Y$ such that
\be
\label{eq:mediatororiginaleqn}
h_{y(z,x)}(g\circ f)_{zx}p_{x}=g_{yz}f_{yx}p_{x}\qquad\forall\;(z,y,x)\in Z\times Y\times X. 
\ee
\end{enumerate}
\en

\bprf
{\color{white}{you found me!}}

\noindent
(\ref{item:atmostone}$\Rightarrow$\ref{item:admitsmediator}) For every $(z,x)\in Z\times(X\setminus N_{p})$ for which such a $y$ exists, set $h(z,x):=y$. If no such $y$ exists or if $x\in N_{p}$, set $h(z,x)$ to be anything. Then $h$ is a mediator for $(f,g)$. 

\noindent
(\ref{item:admitsmediator}$\Rightarrow$\ref{item:mediatorstringexplicit}) Let $h$ be a mediator for $(f,g)$. Since (\ref{eq:mediatororiginaleqn}) holds automatically for $x\in N_{p}$, suppose $x\in X\setminus N_{p}$, in which case (\ref{eq:mediatororiginaleqn}) is equivalent to $h_{y(z,x)}(g\circ f)_{zx}=g_{yz}f_{yx}$ for all $(z,y)\in Z\times Y$. This follows from Equation~(\ref{MFE17}) and the fact that $h$ is a function. 

\noindent
(\ref{item:mediatorstringexplicit}$\Rightarrow$\ref{item:atmostone}) Let $(z,x)\in Z\times (X\setminus N_{p})$ and suppose $(g\circ f)_{zx}>0$. If $h$ is the mediator, then $\sum_{y\in Y}g_{zy}f_{yx}=(g\circ f)_{zx}=g_{zh(z,x)}f_{h(z,x)x}$. But since $g_{zy}f_{yx}=0$ for all $y\ne h(z,x)$, there is only one non-vanishing term in this sum, and it is precisely $g_{zh(z,x)}f_{h(z,x)x}$. 
\eprf

\bt[Functoriality of Conditional Entropy]
\label{thm:deviationK}
Let $(X,p)\xstoch{f}(Y,q)\xstoch{g}(Z,r)$ be a composable pair of morphisms in $\FinPS$. Then 
\be
\label{eqn:conentfun}
H(g\circ f|p)=H(f|p)+H(g|q)
\ee
holds if and only if there exists a mediator $Z\times X\xrightarrow{h}Y$ for $(X,p)\xstoch{f}(Y,q)\xstoch{g}(Z,r)$. 
\et

We first prove two lemmas. 

\blem
\label{lem:conditionalmemoryadds}
Let $(X,p)\xstoch{f}(Y,q)\xstoch{g}(Z,r)$ be a pair of composable morphisms. Then
\[
H\big((g\times\id_{Y})\circ\Delta_{Y}\circ f|p\big)=H(g|q)+H(f|p).
\]
In particular,
$H(g\circ f|p)=H(g|q)+H(f|p)$ if and only if $H(g\circ f|p)=H\big((g\times\id_{Y})\circ\Delta_{Y}\circ f|p\big)$.
\elem

\bprf
[Proof of Lemma~\ref{lem:conditionalmemoryadds}]
On components, 
$\big((g\times\id_{Y})\circ\Delta_{Y}\circ f\big)_{(z,y)x}=g_{zy}f_{yx}$.
Hence,
\[
\begin{split}
H\big((g\times\id_{Y})\circ\Delta_{Y}\circ f|p\big)
&=-\sum_{x}p_{x}\left(\sum_{y,z}g_{zy}f_{yx}\log(g_{zy}f_{yx})\right)\\
&=-\sum_{x,y,z}p_{x}g_{zy}f_{yx}\log(g_{zy})-\sum_{x,y,z}p_{x}g_{zy}f_{yx}\log(f_{yx})\\
&=-\sum_{y,z}q_{y}g_{zy}f_{yx}\log(g_{zy})-\sum_{x,y}p_{x}f_{yx}\log(f_{yx})\\
&=H(g|q)+H(f|p).
\end{split}
\]
Note that this equality still holds if $g_{zy}=0$ or $f_{yx}=0$ as each step in this calculation accounted for such possibilities. 
\eprf

\blem
\label{lem:deviationK}
Let $(X,p)\xstoch{f}(Y,q)\xstoch{g}(Z,r)$ be a pair of composable morphisms in $\FinPS$. Then 
\be
\label{eq:nonnegativeconditionaldifference}
\begin{split}
0&\le H\big((g\times\id_{Y})\circ\Delta_{Y}\circ f|p\big)-H(g\circ f|p)\\
&=-\sum_{x\in X\setminus N_{p}}\sum_{y\in Y\setminus N_{f_{x}}}\sum_{z\in Z\setminus N_{g_{y}}}p_{x}g_{zy}f_{yx}\log\left(\frac{g_{zy}f_{yx}}{\sum_{y'}g_{zy'}f_{y'x}}\right).
\end{split}
\ee
Note that the order of the sums matters in this expression and also note that it is always well-defined since $g_{zy}f_{yx}\ne0$ implies $(g\circ f)_{zx}\ne0$. 
\elem

\bprf
[Proof of Lemma~\ref{lem:deviationK}]
For convenience, temporarily set $\aleph:=H\big((g\times\id_{Y})\circ\Delta_{Y}\circ f|p\big)-H(g\circ f|p)$. Then 
\[
\begin{split}
\aleph
&=
-\sum_{x,y,z}p_{x}g_{zy}f_{yx}\log(g_{zy}f_{yx})+\sum_{x,y,z}p_{x}g_{zy}f_{yx}\log\big((g\circ f)_{zx}\big)\\
&=-\sum_{x\in X\setminus N_{p}}\sum_{y\in Y\setminus N_{f_{x}}}\sum_{z\in Z\setminus N_{g_{y}}}p_{x}g_{zy}f_{yx}\log\left(\frac{g_{zy}f_{yx}}{(g\circ f)_{zx}}\right), 
\end{split}
\]
which proves the claim due to the definition of the composition of stochastic maps. 
\eprf

\bprf
[Proof of Theorem~\ref{thm:deviationK}]
Temporarily set $\aleph:=H\big((g\times\id_{Y})\circ\Delta_{Y}\circ f|p\big)-H(g\circ f|p)$. In addition, note that the set of all $x\in X\setminus N_{p}$ and $z\in Z\setminus N_{(g\circ f)_{x}}$ can be given a more explicit description in terms of the joint distribution $\bullet\xstoch{s:=\gamma\circ\vartheta(g\circ f)}Z\times X$ associated with the composite $g\circ f$ and prior $p$, namely
$s_{(z,x)}:=(g\circ f)_{zx}p_{x}$.
Then,
\be
\label{eq:NpNgfx}
\big\{(z,x)\;:\;x\in X\setminus N_{p},\,z\in Z\setminus N_{(g\circ f)_{x}}\big\}=(Z\times X)\setminus N_{s}. 
\ee

\noindent
($\Rightarrow$) Suppose $\aleph=0$, which is equivalent to Equation~(\ref{eqn:conentfun}) by Lemma~\ref{lem:conditionalmemoryadds}. Then since each term in the sum from Lemma~\ref{lem:deviationK} is non-negative,
\[
0=-g_{zy}f_{yx}p_{x}\log\left(\frac{g_{zy}f_{yx}}{\sum_{y'}g_{zy'}f_{y'x}}\right)
\quad\forall\;x\in X\setminus N_{p},\; y\in Y\setminus N_{f_{x}},\;z\in Z\in N_{g_{y}}.
\]
Hence, fix such an $x\in X\setminus N_{p}, y\in Y\setminus N_{f_{x}},z\in Z\in N_{g_{y}}$. The expression here vanishes if and only if 
\be
\label{eq:gfhasoneterm}
g_{zy}f_{yx}=(g\circ f)_{zx}, \quad\text{ i.e.\ }\quad
g_{zy'}f_{y'x}=0\;\;\forall\;y'\in Y\setminus\{y\}. 
\ee
Hence, for every $x\in X\setminus N_{p}$ and $z\in Z\setminus N_{(g\circ f)_{x}}$, there exists a unique $y\in Y$ such that $g_{zy}f_{yx}\ne 0$.  
But by~(\ref{eq:NpNgfx}), this means that for every $(z,x)\in (Z\times X)\setminus N_{s}$, there exists a unique $y\in Y$ such that $g_{zy}f_{yx}\ne 0$. This defines a function $(Z\times X)\setminus N_{s}\to Y\setminus N_{q}$ which can be extended in an $s$-a.e.\ unique manner to a function $Z\times X\xrightarrow{h}Y.$ 

We now show the function $h$ is in fact a mediator for the composable pair $(g, f)$. 
The equality clearly holds if $x\in N_{p}$ since both sides vanish. Hence, suppose that $x\in X\setminus N_{p}$. Given $y\in Y, z\in Z$, the left-hand-side of~(\ref{eq:mediatororiginaleqn}) equals 
\[
\de_{y h(z,x)}(g\circ f)_{zx}p_{x}
\overset{\text{(\ref{eq:gfhasoneterm})}}{=\joinrel=\joinrel=}
\begin{cases}g_{zy}f_{yx}p_{x}&\mbox{ if $z\in Z\setminus N_{(g\circ f)_{x}}$ and $y=h(z,x)$}\\0&\mbox{ otherwise}\end{cases}.
\]
Similarly, if $x\in X\setminus N_{p}$ and $z\in N_{(g\circ f)_{x}}$, then $g_{zy}f_{yx}=0$ for all $y\in Y$ because otherwise $(g\circ f)_{zx}p_{x}$ would be nonzero. If instead $z\in Z\setminus N_{(g\circ f)_{x}}$, then $g_{zh(z,x)}f_{h(z,x)x}\ne0$ and $g_{zy}f_{yx}=0$ for all $y\in Y\setminus\{h(z,x)\}$ by~(\ref{eq:gfhasoneterm}). Therefore, (\ref{eq:mediatororiginaleqn}) holds. 

\noindent
($\Leftarrow$) Conversely, suppose a mediator $h$ exists and let $X\xstoch{k}Z\times Y$ be the stochastic map given on components by
$k_{(z,y)x}:=h_{y(z,x)}(g\circ f)_{zx}.$ 
Then 
\[
\begin{split}
H(g|q)+H(f|p)&=H\big((g\times\id_{Y})\circ\Delta_{Y}\circ f|p\big)\quad\text{ by Lemma~\ref{lem:conditionalmemoryadds}}\\
&=H(k|p)\quad\hspace{29mm}\text{ by Proposition~\ref{prop:mediatorequivalence} item~(c)}\\
&=-\sum_{x\in X\setminus N_{p}}\sum_{z\in Z\setminus N_{(g\circ f)_{x}}}\sum_{y\in Y}h_{y(z,x)}(g\circ f)_{zx}p_{x}\log\big(h_{y(z,x)}(g\circ f)_{zx}\big)\\
&=-\!\!\!\!\!\!\sum_{(z,x)\in(Z\times X)\setminus N_{s}}\sum_{y\in Y}\de_{yh(z,x)}(g\circ f)_{zx}p_{x}\log\big(\de_{yh(z,x)}(g\circ f)_{zx}\big)\\
&=-\!\!\!\!\sum_{(z,x)\in(Z\times X)\setminus N_{s}}(g\circ f)_{zx}p_{x}\log\big((g\circ f)_{zx}\big)\\
&=H\big((g\circ f)|p\big),
\end{split}
\]
as desired.
\eprf 

\bc[Functoriality of Conditional Information Loss]
\label{FUNCTIL}
Let $(X,p)\xstoch{f}(Y,q)\xstoch{g}(Z,r)$ be a composable pair of morphisms in $\FinPS$. Then $K(g\circ f)=K(f)+K(g)$ if and only if there exists a mediator  $Z\times X\xrightarrow{h}Y$ for the pair $(X,p)\xstoch{f}(Y,q)\xstoch{g}(Z,r)$. 
\ec

\bprf
Since the Shannon entropy difference is always functorial, the conditional information loss is functorial on a pair of morphisms if and only if the conditional entropy is functorial on that pair. Theorem~\ref{thm:deviationK} then completes the proof.  
\eprf

\bx
\label{ex:fdetermimpliescoalescable}
In the notation of Theorem~\ref{thm:deviationK}, suppose that $f$ is \define{a.e.\ deterministic}, which means $f_{yx}=\de_{yf(x)}$ for all $x\in X\setminus N_{p}$ for some function $f$ (abusive notation is used). In this case, the deviation from functoriality,  (\ref{eq:nonnegativeconditionaldifference}), simplifies to 
\[
H\big((g\times\id_{Y})\circ\Delta_{Y}\circ f|p\big)-H(g\circ f|p)
=-\sum_{x\in X\setminus N_{p}}\sum_{z\in Z}p_{x}g_{zf(x)}\log\left(\frac{g_{zf(x)}}{g_{zf(x)}}\right)
=0.
\]
Therefore, if $f$ is  $p$-a.e.\ deterministic, $H(g|q)+H(f|p)=H(g\circ f|p)$. In this case, the mediator $Z\times X\xrightarrow{h}Y$ is given by $h:=!_{Z}\times f$.  
\ex

\bd\label{coalescable}
A pair $(X,p)\xstoch{f}(Y,q)\xstoch{g}(Z,r)$ of composable morphisms in $\FinPS$ is called \define{a.e.\ coalescable} if and only if $(X,p)\xstoch{f}(Y,q)\xstoch{g}(Z,r)$ admits a mediator $Z\times X\xrightarrow{h}Y$. Similarly, a pair $X\xstoch{f}Y\xstoch{g}Z$ of composable morphisms in $\FinStoch$ is called \define{coalescable} iff  $X\xstoch{f}Y\xstoch{g}Z$ admits a strong mediator $Z\times X\xrightarrow{h}Y$. 
\ed

\br
\label{rmk:deterministiccoalescable}
Example~\ref{ex:fdetermimpliescoalescable} showed that if $(X,p)\xstoch{f} (Y,q)$ is $p$-a.e.\ deterministic, then the pair $(X,p)\xstoch{f} (Y,q)\xstoch{g} (Z,r)$ is a.e.\ coalescable for any $g$. In particular, every pair of composable morphisms in $\FinPD$ is coalescable.
\er

In light of Theorem~\ref{thm:deviationK} and Corollary~\ref{FUNCTIL}, we make the following definition, which will serve as one of the axioms in our later characterizions of both conditional information loss and conditional entropy.   

\bd\label{sfunct}
A function $F:\FinPS\to \B\R$ is said to be \define{semi-functorial} iff $F(g\circ f)=F(g)+F(f)$ for every a.e.\ coalescable pair $(X,p)\xstoch{f}(Y,q)\xstoch{g}(Z,r)$ in $\FinPS$.
\ed 

\bx
By Theorem~\ref{thm:deviationK} and Corollary~\ref{FUNCTIL}, conditional information loss and conditional entropy are both semi-functorial.
\ex

\bn\label{frestrict}
Suppose $F:\FinPS\to \B\R$ is semi-functorial. Then the restriction of $F$ to $\FinPD$ is functorial.  
In particular, if $F$ is, in addition, convex linear, continuous, and reductive, then $F$ is a non-negative multiple of conditional information loss.
\en

\bprf
By Example~\ref{ex:fdetermimpliescoalescable}, every pair of composable morphisms in $\FinPD$ is a.e.\ coalescable. Therefore, $F$ is functorial on $\FinPD$. The second claim then follows from Proposition~\ref{prop:prelimcharacterization}.
\eprf

The following lemma will be used in later sections and serves to illustrate some examples of a.e.\ coalescable pairs. 

\blem
\label{lem:wedgingwithisos}
Let $(W,s)\xrightarrow{e}(X,p)\xstoch{f}(Y,q)\xrightarrow{g}(Z,r)$ be a triple of composable morphisms with $e$ deterministic and $g$ invertible. Then each of the following pairs are a.e.\ coalescable: 
\begin{enumerate}[i)]
\itemsep0pt
\item
$(W,s)\xrightarrow{e}(X,p)\xstoch{f}(Y,q)$
\item
$(X,p)\xstoch{f}(Y,q)\xrightarrow{g}(Z,r)$
\item
$(W,s)\xrightarrow{e}(X,p)\xstoch{g\circ f}(Z,r)$
\item
$(W,s)\xstoch{f\circ e}(Y,q)\xrightarrow{g}(Z,r)$
\end{enumerate}
\elem

\bprf
The proof that $(W,s)\xrightarrow{e}(X,p)\xstoch{f}(Y,q)$ is coalescable was provided (in a stronger form) in Example~\ref{ex:fdetermimpliescoalescable}. To see that $(X,p)\xstoch{f}(Y,q)\xrightarrow{g}(Z,r)$ is coalescable, note that since $g$ is an isomorphism we have $(g\circ f)_{zx}=g_{zg^{-1}(z)}f_{g^{-1}(z)x}$.
Thus, $Z\times X\xrightarrow{g^{-1}\times !_{X}}Y\times\bullet\cong Y$ is a mediator function for $g\circ f$, thus $g\circ f$ is coalescable. The last two claims follow from the proofs of the first two claims.
\eprf

\section{Bayesian inversion}
\label{sec:Bayesianinversion}

In this section, we recall the concepts of a.e.\ equivalence and Bayesian inversion phrased in a categorical manner~\cites{Fo12,ChJa18,Fr19}, as they will play a significant role moving forward. 

\bd
Let $(X,p)\xstoch{f}(Y,q)$ and $(X,p)\xstoch{g}(Y,q)$ be two morphisms in $\FinPS$ with the same source and target. Then $f$ and $g$ are said to \define{almost everywhere equivalent} (or $p$-$a.e.$ \define{equivalent}) if and only if  $f_{yx}=g_{yx}$ for every $x\in X$ with $p_x\neq 0$. In such a case, the $p$-$a.e.$ equivalence of $f$ and $g$ will be denoted $f\underset{\raisebox{.6ex}[0pt][0pt]{\scriptsize$p$}}{=}g$. 
\ed


\bt[Bayesian Inversion~\cites{Fo12,ChJa18,PaBayes}]
\label{thm:classicalBayestheorem}
Let $(X,p)\xstoch{f}(Y,q)$ be a morphism in $\FinPS$. Then there exists a morphism $(Y,q)\xstoch{\ov f}(X,p)$ 
such that $\ov f_{xy}q_{y}=f_{yx}p_{x}$ for all $x\in X$ and $y\in Y$. 
Furthermore, for any other morphism $(Y,q)\xstoch{\ov f'}(X,p)$ satisfying this condition, $\ov f\underset{\raisebox{.6ex}[0pt][0pt]{\scriptsize$q$}}{=}\ov f'$. 
\et

\bd\label{RBIF71}
The morphism $(Y,q)\xstoch{\ov f}(X,p)$ appearing in Theorem~\ref{thm:classicalBayestheorem} will be referred to as a  \define{Bayesian inverse} of $(X,p)\xstoch{f} (Y,q)$. It follows that $\overline{f}_{xy}=p_xf_{yx}/q_y$ for all $y \in Y$ with $q_y\neq 0$.
\ed

\bn
\label{lem:Binversionproperties}
Bayesian inversion satisfies the following properties. 
\begin{enumerate}[i)]
\itemsep0pt
\item
\label{BIPAEQ19}
Suppose $(X,p)\xstoch{f}(Y,q)$ and $(X,p)\xstoch{g}(Y,q)$ are $p$-a.e. equivalent, and let $\ov f$ and $\ov g$ be Bayesian inverses of $f$ and $g$ respectfully. Then $\ov f\underset{\raisebox{.6ex}[0pt][0pt]{\scriptsize$q$}}{=}\ov g$.
\item
\label{item:binvsymm}
Given two morphisms $(X,p)\xstoch{f}(Y,q)$ and $(Y,q)\xstoch{g}(X,p)$ in $\FinPS$, then $f$ is a Bayesian inverse of $g$ if and only if $g$ is a Bayesian inverse of $f$. 
\item
\label{ISOBI19}
Let $(Y,q)\xstoch{\ov f} (X,p)$ be a Bayesian inverse of $(X,p)\xstoch{f} (Y,q)$, and let $\gamma:X\times Y\to Y\times X$ be the swap map (as in Definition~\ref{DSM77}).  Then $\vartheta(f)=\gamma\circ \vartheta(\ov f)$
\item
\label{CBI997}
Let $(f,g)$ be a composable pair of morphisms in $\FinPS$, and suppose $\ov f$ and $\ov g$ are Bayesian inverses of $f$ and $g$ respectively. Then $(\ov g, \ov f)$ is a composable pair, and $\ov f\circ \ov g$ is a Bayesian inverse of $g\circ f$. 
\end{enumerate}
\en

\bprf
These are immediate consequences of the categorical definition of a Bayesian inverse (see~\cites{Fr19,Sm20,PaBayes} for proofs).
\eprf

\bd\label{BIFD71}
A contravariant function $\mathcal{B}:\FinPS\to \FinPS$ is said to be a \define{Bayesian inversion functor}%
\footnote{This is mildly abusive terminology since functoriality only holds in the a.e.\ sense (cf.\ Remark~\ref{rmk:Bayesianinversionfunctor}).}
 if and only if $\mathcal{B}$ acts as the identity on objects and $\mathcal{B}(f)$ is a Bayesian inverse of $f$ for all morphisms $f$ in $\FinPS$.
\ed

\br
\label{rmk:Bayesianinversionfunctor}
A Bayesian inversion functor exists. Given any $(X,p)\xstoch{f}(Y,q)$, set $Y\xstoch{\overline{f}}X$ to be given by $\overline{f}_{xy}=p_xf_{yx}/q_y$ for all $y \in Y$ with $q_y\neq 0$ and $\overline{f}_{xy}=1/|X|$ for all $y\in Y$ with $q_{y}=0$. Note that this does \emph{not} define a functor. Indeed, if $(X,p)$ is a probability space with $p_{x_{0}}=0$ for some $x_{0}\in X$, then $(\overline{\id_{X}})_{x_{0}}$ is the uniform measure on $X$ instead of the Dirac delta measure concentrated on $x_{0}$. In other words, $\overline{\id_{X}}\ne\id_{X}$. Similar issues of measure zero occur, indicating that $\overline{g\circ f}\ne \ov f\circ \ov g$ for a composable pair of morphisms $(X,p)\xstoch{f}(Y,q)\xstoch{g}(Z,r)$. Nevertheless, Bayesian inversion is a.e.\ functorial in the sense that $\overline{g\circ f}\aeequals{r}\ov f\circ\ov g$ and $\overline{\id_{(X,p)}}\aeequals{p}\id_{(X,p)}$. 
\er

\begin{corollary}
\label{cor:B2equalsid}
$\mathcal{B}^2(f)\aeequals{p}f$ for any Bayesian inversion functor $\mathcal{B}$ and every $(X,p)\xstoch{f}(Y,q)$ in $\FinPS$.
\end{corollary}

\bn
\label{BICC79}
Let $\mathcal{B}$ be a Bayesian inversion functor on $\FinPS$ (as in Definition~\ref{BIFD71}). Then $\mathcal{B}$ is \define{a.e.\  convex linear} in the sense that 
\[
\mathcal{B}\left(\bigoplus_{x\in X}p_xQ^x\right)\aeequals{q'}\bigoplus_{x\in X}p_x\mathcal{B}(Q^x),
\]
where $q':=\bigoplus_{x\in X}p_{x}q'^{x}$ and the other notation is as in Definition~\ref{defn:convexcombos}. 
\en

\bprf
First note that it is immediate that $\mathcal{B}$ is convex linear on objects since Bayesian inversion acts as the identity on objects. 
Let $p:\bullet\xstoch{} X$ be a probability measure, $(Y_x,q^x)\xstoch{Q^x} (Y'_x,q'^x)$ be a collection of morphisms in $\mathbf{FinPS}$ indexed by $X$, and suppose $\mathcal{B}$ is a Bayesian inversion functor. Then for $(z,z')\in Y_x\times Y'_x$ with $p_xq'^x_{z'}\neq 0$, we have
\begin{eqnarray*}
\mathcal{B}\left(\bigoplus_{x\in X}p_xQ^x\right)_{zz'}&=&\frac{p_xq^x_z\left(\bigoplus_{x\in X}p_xQ^x\right)_{z'z}}{p_xq'^x_{z'}} =\frac{q^x_z\left(Q^x_{z'z}\right)}{q'^x_{z'}} 
\\
&=&\mathcal{B}(Q^x)_{zz'}=\left(\bigoplus_{x\in X}p_x\mathcal{B}(Q^x)\right)_{zz'}. \\
\end{eqnarray*}
Thus, $\mathcal{B}$ is a.e.\  convex linear.
\eprf


\bn\label{BICP979}
Given $(X,p)\xstoch{f} (Y,q)\xstoch{g} (Z,r)$ in $\FinPS$, and let $\ov f$ and $\ov g$ be Bayesian inverses of $f$ and $g$ respectively. Then $(f,g)$ is a.e.\ coalescable if and only if $(\ov g, \ov f)$ is a.e.\ coalescable. 
\en

\bprf
Since Bayesian inversion is a dagger functor on a.e.\ equivalence classes~\cite[Remark~13.9]{Fr19}, it suffices to prove one direction in this claim. Hence, suppose $(f,g)$ is a.e.\ coalescable and let $h$ be a mediator function realizing this. Then $h\circ\gamma$ is a mediator for $(\ov g, \ov f)$ because
\[
\begin{split}
\ov f_{xy}\ov g_{yz} r_{z}&=\ov f_{xy}g_{zy}q_{y}=g_{zy}f_{yx}p_{x}=h_{y(z,x)}(g\circ f)_{zx}p_{x}\\
&=h_{y(z,x)}\overline{(g\circ f)}_{xz}r_{z}=(h\circ\gamma)_{y(x,z)}\overline{(g\circ f)}_{xz}r_{z}. 
\end{split}
\]
A completely string-diagrammatic proof is provided in Appendix~\ref{app:markovcat}. 
\eprf

The following proposition is a reformulation of the conditional entropy identity $H(\mathcal{Y}|\mathcal{X})+H(\mathcal{X})=H(\mathcal{X}|\mathcal{Y})+H(\mathcal{Y})$ in terms of Bayesian inversion.

\bn
\label{EBRCE}
Let $(X,p)\xstoch{f} (Y,q)$ be a morphism in $\FinPS$, and suppose $\ov f$ is a Bayesian inverse of $f$. Then
\be\label{CEI19}
H(f|p)+H(p)=H(\ov f|q)+H(q).
\ee 
\en 

\bprf
This follows from the fact that both sides of \eqref{CEI19} are equal to $H(\vartheta(f))$. 
\eprf

Proposition~\ref{EBRCE} implies Bayesian inversion takes conditional entropy to conditional information loss and vice versa, which is formally stated as follows.

\bc
\label{prop:Bayesswaps}
Let $K:\FinPS\to\B\R_{\ge0}$ and $H:\FinPS\to\B\R_{\ge0}$ be given by conditional information loss and conditional entropy, respectively, and let $\mathcal{B}:\FinPS\to\FinPS$ be a Bayesian inversion functor. Then%
\footnote{Since contravariant functions into $\B\R_{\ge0}$ are covariant and vice versa (cf.\ Convention~\ref{conv:functions}), the composites $K\circ\mathcal{B}$ and $H\circ\mathcal{B}$ are all covariant, which is why we have not written any ${}^{\op}$ superscripts.}
$H=K\circ\mathcal{B}$ and $K=H\circ\mathcal{B}$.
\ec

\br
If $(X,p)\overset{f}\to (Y,q)$ is a deterministic morphism in $\FinPS$, Baez, Fritz, and Leinster point out that the information loss of $f$ is in fact the conditional entropy of $x$ given $y$~\cite{BFL}.
Here, we see this duality as a special case of Corollary~\ref{prop:Bayesswaps} applied to deterministic morphisms.
\er

\section{Bloom-shriek factorization}

We now introduce a simple, but surprisingly useful, factorization for every morphism in $\FinPS$, and we use it to prove some essential lemmas for our characterization theorems for conditional information loss and conditional entropy, which appear in the following sections.  

\bd
\label{defn:bloomshriek}
Given a stochastic map $X\xstoch{f}Y$, the \define{bloom of $f$} is the stochastic map $X\xstoch{\bloom_{f}}X\times Y$ given by the composite $X\xstoch{\Delta_{X}}X\times X\xstoch{\id_{X}\times f}X\times Y$, and the \define{shriek of $f$} is the deterministic map $X\times Y\xrightarrow{!_{f}}X$ given by the projection $\pi_{X}$. 
\ed

\bn
\label{prop:bloomshriekfactorization}
Let $(X,p)\xstoch{f}(Y,q)$ be a morphism in $\FinPS$. Then the following statement hold. 
\begin{enumerate}[i)]
\itemsep0pt
\item
The composite $(X,p)\xstoch{\bloom_{f}} (X\times Y,\vartheta(f))\xrightarrow{\shriek_{f}} (X,p)$ is equal to the identity $\emph{id}_{X}$. 
\item
\label{BSF17}
The morphism $f$ equals the composite $(X,p)\xstoch{\bloom_{f}}(X\times Y,\vartheta(f))\xrightarrow{!_{\ov f}\circ\gamma}(Y,q)$, where $\ov f$ denotes any Bayesian inverse of $f$ and $\gamma:X\times Y\to Y\times X$ is the swap map. 
\item\label{BSCOAL19}
The pair $(X,p)\xstoch{\bloom_{f}}(X\times Y,\vartheta(f))\xrightarrow{!_{\ov f}\circ\gamma\equiv\pi_{Y}}(Y,q)$ is coalescable.
\end{enumerate} 
\en

\bd
The decomposition in item \ref{BSF17} Proposition~\ref{prop:bloomshriekfactorization} will be referred to as the \define{bloom-shriek factorization} of $f$. 
\ed

\bprf
[Proof of Propostion~\ref{prop:bloomshriekfactorization}]
Element-wise proofs are left as exercises. Appendix~\ref{app:markovcat} contains an abstract proof using string diagrams in Markov categories.  
\eprf

The bloom of $f$ can be expressed as a convex combination of simpler morphisms up to isomorphism. To describe this and its behavior under convex linear semi-functors, we introduce the notion of an invariant and examine some of its properties.

\bd
\label{defn:invariant}
A function $F:\FinPS\to \B\R$ is said to be an \define{invariant} if and only if for every triple of composable morphisms $(W,s)\xrightarrow{e}(X,p)\xstoch{f}(Y,q)\xrightarrow{g}(Z,r)$ such that $e$ and $g$ are isomorphisms, then  $F(f)=F(g\circ f\circ e)$. 
\ed


\blem
\label{lem:Fsemifunctinvariant}
If a function $F:\FinPS\to\B\R_{\ge0}$ is semi-functorial, then $F$ is an invariant.
\elem

\bprf
Consider a composable triple $(W,s)\xrightarrow{e}(X,p)\xstoch{f}(Y,q)\xrightarrow{g}(Z,r)$ such that $e$ and $g$ are isomorphisms. Then
\[
F\big(g\circ f\circ e\big)
=F\big(g\circ (f\circ e)\big)
=F(g)+F(f\circ e)
=F(g)+F(f)+F(e)
\]
by Lemma~\ref{lem:wedgingwithisos}. Secondly, since $g$ and $e$ are isomorphisms, and since the pairs $(g,g^{-1})$ and $(e,e^{-1})$ are coalescable, $F(\id)=F(g\circ g^{-1})=F(g)+F(g^{-1})$. But since $F(\id)=0$ (by semi-functoriality), this requires that $F(g)=0$ for an isomorphism $g$ since $F(g)\ge0$ and $F(g^{-1})\ge0$. The same is true for $F(e)$. Hence, $F\big(g\circ f\circ e\big)=F(f)$. 
\eprf

\blem
\label{lem:decomposeFfbloomshriek}
Let $(X,p)\xstoch{f}(Y,q)$ be a morphism in $\FinPS$, and suppose $F:\FinPS\to\B\R_{\geq 0}$ is semi-functorial and convex linear. Then the following statements hold.
\begin{enumerate}[i)] 
\itemsep0pt
\item\label{LSX1}
$F(f)=F(\shriek_{\ov f})+F(\bloom_{f})$
\item\label{LSX2}
$F(\bloom_f)=\sum_{x\in X}p_xF(\bloom_{f_x})$
\item\label{LSX3}
$F(\shriek_f)=\sum_{x\in X}p_xF(\shriek_{f_x})$
\end{enumerate}
\elem

\bprf
For item \ref{LSX1}, we have
\begin{align*}
F(f)&=F(!_{\ov f}\circ \gamma \circ \bloom_f)&&\text{ by item~\ref{BSF17} of Proposition~\ref{prop:bloomshriekfactorization}}\\
&=F(!_{\ov f}\circ \gamma)+F(\bloom_f)&&\text{ by item \ref{BSCOAL19} of Proposition~\ref{prop:bloomshriekfactorization}}\\
&=F(\shriek_{\ov f})+F(\bloom_{f})&&\text{ by Lemma~\ref{lem:Fsemifunctinvariant}}.
\end{align*}

For items \ref{LSX2} and \ref{LSX3}, note that $\bloom_{f}$ and $\shriek_{f}$ can be expressed as composites of isomorphisms and certain convex combinations, namely
\be
\label{eq:decomposebloomf}
\xy0;/r.25pc/:
(-22.5,7.5)*+{(X,p)}="X";
(-22.5,-7.5)*+{\ds\bigoplus_{x\in X}p_{x}(\bullet,1)}="px";
(22.5,-7.5)*+{\ds\bigoplus_{x\in X}p_{x}(Y,f_{x})}="fx";
(22.5,7.5)*+{\big(X\times Y,\vartheta(f)\big)}="XY";
{\ar"X";"px"_(0.4){\cong}^(0.4){\psi}};
{\ar"fx";"XY"_(0.55){\cong}^(0.55){\varphi}};
{\ar@{~>}"X";"XY"^(0.4){\bloom_{f}}};
{\ar@{~>}"px";"fx"_{\ell:={\text{\footnotesize$\ds\bigoplus_{x\in X}p_{x}\bloom_{f_{x}}$}}}};
\endxy
\;\text{ and }\;
\xy0;/r.25pc/:
(-22.5,7.5)*+{\big(X\times Y,\vartheta(f)\big)}="X";
(-22.5,-7.5)*+{\ds\bigoplus_{x\in X}p_{x}(Y,f_x)}="px";
(22.5,-7.5)*+{\ds\bigoplus_{x\in X}p_{x}(\bullet,1)}="fx";
(22.5,7.5)*+{(X,p)}="XY";
{\ar"X";"px"_(0.4){\cong}^(0.4){\psi'}};
{\ar"fx";"XY"_(0.55){\cong}^(0.55){\varphi'}};
{\ar@{~>}"X";"XY"^(0.55){\shriek_{f}}};
{\ar@{~>}"px";"fx"_{\ell':={\text{\footnotesize$\ds\bigoplus_{x\in X}p_{x}\shriek_{f_{x}}$}}}};
\endxy
\;.
\ee
Hence,
\begin{align*}
F(\bloom_f)&=F(\varphi\circ\ell\circ\psi)&F(\shriek_{f})&=F(\varphi'\circ\ell'\circ\psi')&&\text{ by (\ref{eq:decomposebloomf})}\\
&=F(\ell)&&=F(\ell')&&\text{ by Lemma~\ref{lem:Fsemifunctinvariant}}\\
&=\sum_{x\in X}p_{x}F(\bloom_{f_{x}})&&=\sum_{x\in X}p_{x}F(\shriek_{f_{x}})&&\text{ since $F$ is convex linear.}
\qedhere
\end{align*} 
\eprf

\bn
\label{P717}
Suppose $F:\FinPS\to\B\R_{\geq 0}$ is semi-functorial and convex linear. 
If $f,g:(X,p)\stoch(Y,q)$ are two morphisms in $\FinPS$ such that $f\aeequals{p}g$, then $F(f)=F(g)$. 
\en

\bprf
Suppose $(X,p)\xstoch{f} (Y,q)$ and $(X,p)\xstoch{g} (Y,q)$ are such that $f\underset{\raisebox{.6ex}[0pt][0pt]{\scriptsize$p$}}{=}g$, and let $\ov f$ and $\ov g$ be Bayesian inverses for $f$ and $g$. 
Then 
\begin{align*}
F(f)&=F(\shriek_{\ov f})+F(\bloom_{f})&&\text{ by item \ref{LSX1} of Lemma~\ref{lem:decomposeFfbloomshriek}}\\
&=\sum_{y\in Y}q_{y}F(\shriek_{\ov f_{y}})+\sum_{x\in X}p_{x}F(\bloom_{f_{x}})&&\text{ by items \ref{LSX2} and \ref{LSX3} of Lemma~\ref{lem:decomposeFfbloomshriek}}\\
&=\sum_{y\in Y}q_{y}F(\shriek_{\ov g_{y}})+\sum_{x\in X}p_{x}F(\bloom_{g_{x}})&&\text{ since $f\aeequals{p}g$ and $\ov f\aeequals{q}\ov g$}\\
&=F(\shriek_{\ov g})+F(\bloom_{g})&&\text{ by items \ref{LSX2} and \ref{LSX3} of Lemma~\ref{lem:decomposeFfbloomshriek}}\\
&=F(g)&&\text{ by item \ref{LSX1} of Lemma~\ref{lem:decomposeFfbloomshriek},}\\ 
\end{align*}
as desired.
\eprf

\section{An intrinsic characterization of conditional information loss}\label{CT771}

\bt\label{CTKIL89}
Suppose $F:\FinPS\to \B\R_{\geq 0}$ is a function satisfying the following conditions.
\begin{enumerate}
\itemsep0pt
\item\label{K1}
$F$ is semi-functorial.
\item\label{K2}
$F$ is convex linear.
\item\label{K3}
$F$ is continuous.
\item\label{K4}
$F(\bloom_p)=0$ for every probability distribution $\bullet\xstoch{p} X$.
\end{enumerate}
Then $F$ is a non-negative multiple of conditional information loss. Conversely, conditional information loss satisfies conditions \ref{K1}-\ref{K4}.
\et

\bprf
Suppose $F$ satisfies conditions \ref{K1}-\ref{K4}, let $(X,p)\xstoch{f} (Y,q)$ be an arbitrary morphism in $\FinPS$, and let $\gamma:Y\times X\to X\times Y$ be the swap map, so that $\shriek_{\ov f}=\pi_Y\circ \gamma$. Then
\begin{align*}
F(f)&=F(\shriek_{\ov f})+F(\bloom_{f})&&\text{ by item \ref{LSX1} of Lemma~\ref{lem:decomposeFfbloomshriek}}\\
&=F(\shriek_{\ov f})+\sum_{x\in X}p_xF(\bloom_{f_x})&&\text{ by item \ref{LSX2} of Lemma~\ref{lem:decomposeFfbloomshriek}}\\
&=F(\shriek_{\ov f})&&\text{ by condition \ref{K4}} \\
&=F\left(\pi_Y\circ \gamma\right) \\
&=F(\pi_Y) && \text{ since $F$ is an invariant (Definition~\ref{defn:invariant}).}
\end{align*}
Thus, $F$ is reductive (see Definition~\ref{redux89}) 
and Proposition~\ref{frestrict} applies.
\eprf

\br
Under the assumption that $F:\FinPS\to \B\R$ is semi-functorial and convex linear, one may show $F$ satisfies condition \ref{K4} in Theorem~\ref{CTKIL89} if and only if $F$ is reductive (see Definition~\ref{redux89} and Proposition~\ref{frestrict}). 
While the reductive axiom specifies how the semi-functor acts on \emph{all} morphisms in $\FinPS$, condition~\ref{K4} in Theorem~\ref{CTKIL89} only specifies how it acts on morphisms from the \emph{initial} object. This gives not just a simple mathematical criterion, but one with a simple intuitive interpretation as well.  
Namely, condition \ref{K4} says that if a process begins with no prior information, then there is no information to be lost in the process.
\er

We now use Theorem~\ref{CTKIL89} and Bayesian inversion to prove a statement dual to Theorem~\ref{CTKIL89}. 

\bt\label{CTKIL97}
Suppose $F:\FinPS\to \B\R_{\geq 0}$ is a function satisfying the following conditions.
\begin{enumerate}
\itemsep0pt
\item\label{S1}
$F$ is semi-functorial.
\item\label{S2}
$F$ is convex linear.
\item\label{S3}
$F$ is continuous.
\item\label{S4}
$F(\shriek_p)=0$ for every probability distribution $\bullet\xstoch{p} X$.
\end{enumerate}
Then $F$ is a non-negative multiple of conditional entropy. Conversely, conditional entropy satisfies conditions \ref{S1}-\ref{S4}.
\et

Before giving a proof, we introduce some terminology and prove a few lemmas. We also would like to point out that condition \ref{S4} may be given an operational interpretation as follows: if a communication channel has a constant output, then it has no conditional entropy.

\bd
Let $F:\FinPS\to \B\R$ be a function and let $\mathcal{B}$ be a Bayesian inversion functor. Then $\ov F:=F\circ \mathcal{B}$ will be referred to as a \define{Bayesian reflection} of $F$.
\ed

\br
By Proposition~\ref{P717}, if $F:\FinPS\to \B\R$ is a convex linear semi-functor, then a  Bayesian reflection is independent of the choice of a Bayesian inversion functor, and as such, is necessarily unique. 
\er

\blem\label{reflex}
Let $(X,p)\xstoch{f} (Y,q)$ be a morphism in $\FinPS$, suppose $F:\FinPS\to \B\R$ is a convex linear semi-functor, and let $\ov f$ be a Bayesian inverse of $f$. Then $\ov F(f)=F(\ov f)$.
\elem

\bprf
[Proof of Lemma~\ref{reflex}]
Let $\mathcal{B}$ be a Bayesian inversion functor, so that $\mathcal{B}(f)\aeequals{q} \ov f$. Then
$
\ov F(f)=F(\mathcal{B}(f))=F(\ov f),
$
where the last equality follows from Proposition~\ref{P717}.
\eprf

\blem
\label{BIAEC}
Let $\mathcal{B}:\FinPS\to\FinPS$ be a Bayesian inversion functor and let $(X_{n},p_{n})\xstoch{f_{n}}(Y_{n},q_{n})$ be a sequence of morphisms in $\FinPS$ converging to $(X,p)\xstoch{f}(Y,q)$. Then $\ds\lim_{n\to\infty}\mathcal{B}(f_{n})\aeequals{q}\mathcal{B}(f)$. 
\elem

\bprf
[Proof of Lemma~\ref{BIAEC}]
Set $f^{(n)}:=f_{n}$. 
For all $y\in Y$ with $q_y\neq 0$, we have
\[
\lim_{n\to\infty}\mathcal{B}\big(f^{(n)}\big)_{xy}=\lim_{n\to\infty}\frac{p^{(n)}_{x}f^{(n)}_{yx}}{q^{(n)}_y}=\frac{p_xf_{yx}}{q_y}=\mathcal{B}(f)_{xy}. \qedhere
\]
\eprf

\blem
\label{lem:iffbayes}
Suppose $F:\FinPS\to\B\R_{\ge0}$ is a function satisfying conditions \ref{S1}-\ref{S4} of Theorem~\ref{CTKIL97}. Then the Bayesian reflection $\ov F$ is a non-negative multiple of conditional information loss.
\elem

\bprf
[Proof of Lemma~\ref{lem:iffbayes}]
We show $\ov F$ satisfies conditions \ref{K1}-\ref{K4} of Theorem~\ref{CTKIL89}. Throughout the proof, let $\mathcal{B}$ denote a Bayesian inversion functor, so that $\ov F=F\circ \mathcal{B}$.

\noindent
\underline{\smash{Semi-functoriality}}: Suppose $(X,p)\xstoch{f} (Y,q)\xstoch{g} (Z,r)$ is an a.e.\ coalescable pair of composable morphisms in $\FinPS$. Then
\begin{align*}
\ov F(g\circ f)&=F\big(\ov {g\circ f}\big)&&\text{ by Lemma~\ref{reflex} }\\
&=F\big(\ov f\circ\ov g\big)&&\text{ by item \ref{CBI997} of Proposition~\ref{lem:Binversionproperties} }\\
&=F(\ov f)+F(\ov g)&&\text{ by Proposition~\ref{BICP979}}\\
&=\ov F(f)+\ov F(g)&&\text{ by Lemma~\ref{reflex}.}
\end{align*}
Thus, $\ov F$ is semi-functorial. 

\noindent
\underline{\smash{Convex Linearity}}: Given any probability space $(X,p)$ and a family of morphisms $(Y_x,q^x)\xstoch{Q^x} (Y'_x,q'^x)$ in $\mathbf{FinPS}$ indexed by $X$, 
\begin{align*}
\ov F\left(\bigoplus_{x\in X}p_{x}Q^{x}\right)&=F\left(\bigoplus_{x\in X}p_{x}\mathcal{B}(Q^{x})\right)&&\text{ by Propositions~\ref{BICC79} and~\ref{P717}}\\
&=\sum_{x\in X}p_{x}F\big(\mathcal{B}(Q^{x})\big)&&\text{ since $F$ is convex linear}\\
&=\sum_{x\in X}\ov F(Q^{x})&&\text{ by definition of $\ov F$.}
\end{align*}
Thus, $\ov F$ is convex linear.  

\noindent
\underline{\smash{Continuity}}: This follows from Lemma~\ref{BIAEC} and Proposition~\ref{P717}.

\noindent
\underline{$\ov F(\bloom_p)=0$ for every probability distribution $\bullet\xstoch{p} X$}: This follows from Lemma~\ref{reflex}, since $\shriek_p$ is the unique Bayesian inverse of $\bloom_p$.
\eprf

\bprf[Proof of Theorem~\ref{CTKIL97}]
Suppose $F:\FinPS\to \B\R$ is a function satisfying conditions \ref{S1}-\ref{S4} of Theorem~\ref{CTKIL97}, and let $\mathcal{B}$ be a Bayesian inversion functor. Since $F$ is semi-functorial and convex linear it follows from Proposition~\ref{P717} that $F=\ov F\circ \mathcal{B}$, and by Lemma~\ref{lem:iffbayes} it follows that $\ov F=cK$ for some non-negative constant $c\geq 0$. We then have $F=\ov F\circ \mathcal{B}=cK\circ \mathcal{B}=cH$, thus $F$ is a non-negative multiple of conditional entropy.
\eprf

\section{A Bayesian characterization of conditional entropy}\label{entropicbayes}

We now prove a reformulation of Theorem~\ref{CTKIL97} where condition \ref{S4} is replaced by a condition which we view as an `entropic Bayes' rule'. 

\bd
\label{entropix}
A function $F:\FinPS\to \B\R$ satisfies an \define{entropic Bayes' rule} if and only if 
\[
F(f)+F(\bloom_p)=F(\ov f)+F(\bloom_q)
\]
for every morphism $(X,p)\xstoch{f} (Y,q)$ in $\FinPS$ and any Bayesian inverse $\ov f$ of $f$.
\ed

\br
The entropic Bayes' rule is an abstraction of the conditional entropy identity \eqref{CEI19}.
\er 

\bt[A Bayesian characterization of conditional entropy]
\label{BCE19}
Suppose $F:\FinPS\to \B\R_{\geq 0}$ is a function satisfying the following conditions.
\begin{enumerate}
\itemsep0pt
\item\label{BC1}
$F$ is semi-functorial.
\item\label{BC2}
$F$ is convex linear.
\item\label{BC3}
$F$ is continuous.
\item\label{BC4}
$F$ satisfies an entropic Bayes' rule.
\end{enumerate}
Then $F$ is a non-negative multiple of conditional entropy. Conversely, conditional entropy satisfies conditions \ref{S1}-\ref{S4}.
\et

\bprf
By Theorem~\ref{CTKIL97}, it suffices to show $F(\shriek_p)=0$ for every object $(X,p)$ in $\FinPS$. 
For this, first note that $\bloom_1=\bloom_1\circ \bloom_1$, where $(\bullet,1)$ is the point-mass distribution on a single point. Since $F$ is semi-functorial and $\bloom_1\circ \bloom_1$ is coalescable, we have
$F(\bloom_1)=F(\bloom_1)+F(\bloom_1)$, which implies $F(\bloom_1)=0$. 
Applying the entropic Bayes' rule from Definition~\ref{entropix}  
to the morphism $\shriek_p:(X,p)\xstoch{} (\bullet,1)$
yields
\[
F(\shriek_p)+F(\bloom_p)=F(\bloom_p)+F(\bloom_1)\implies F(\shriek_p)=F(\bloom_1)=0,
\] 
as desired.
\eprf

\br
In~\cite[slide 21]{Fr20}, Fritz asked if there is a Markov category for information theory explaining the analogy between the Bayes rule $\mathbb{P}(A|B)\mathbb{P}(B)=\mathbb{P}(B|A)\mathbb{P}(A)$ and the conditional entropy identity $H(A|B)+H(B)=H(B|A)+H(A)$. In light of our work, we feel we have an adequate categorical explanation for this analogy, which we now explain. 

Let $(X,p)\xstoch{f} (Y,q)$ be an arbitrary morphism in $\FinPS$, and suppose $F:\FinPS\to \B\R$ is semi-functorial. Then the commutative diagram (cf.\ Definition~\ref{defn:bayesinversestring})
\be
\label{bayesquare}
\xy0;/r.22pc/:
(0,10)*+{(\bullet,1)}="X";
(-28,0)*+{(X,p)}="px";
(0,-10)*+{(X\times Y,\vartheta(f))}="fx";
(28,0)*+{(Y,q)}="XY";
{\ar@{~>}"X";"px"_{\bloom_p}};
{\ar@{~>}"XY";"fx"^(0.4){\gamma\circ  \bloom_{\ov f}}};
{\ar@{~>}"X";"XY"^{\bloom_{q}}};
{\ar@{~>}"px";"fx"_(0.4){\bloom_f}};
\endxy
\ee
is a coalescable square (where $\gamma$ is the swap map), i.e., $\bloom_f\circ \bloom_p$ and $(\gamma\circ \bloom_{\ov f})\circ \bloom_q$ are both coalescable. The semi-functoriality of $F$ then implies the identity $F(\bloom_f)+F(\bloom_p)=F(\bloom_{\ov f})+F(\bloom_q)$. Now suppose---as in the case of conditional entropy---that $F$ satisfies the further condition that $F(f)=F(\bloom_f)$. Then commutivity of \eqref{bayesquare} and this 
are equivalent to the following two respective equations: 
\begin{align*}
\text{\underline{\smash{Bayes' Rule}}}&: \quad 
 &f_{yx} p_{x}&=\ov f_{xy} q_{y}\\
\underline{\text{\smash{Entropic Bayes' Rule}}}&: \quad
&F(f)+F(\bloom_p)&=F(\ov f)+F(\bloom_q).
\end{align*}
In the case that $F=H$, where $H$ is the conditional entropy, we have $H(\bloom_r|1)=H(r)$ for every object $(Z,r)$ in $\FinPS$ (where $H(r)$ is the Shannon entropy). Thus, the entropic Bayes' rule becomes $H(f|p)+H(p)=H(\ov f|q)+H(q)$, which is the classical identity for conditional entropy.
\er

\appendix
\section{Correctable codes and conditional information loss}
\label{app:cccil}

In this appendix, we prove that the conditional information loss of a morphism $(X,p)\xstoch{f}(Y,q)$ in $\FinPS$ vanishes if and only if $f$ is a disintegration, or equivalently, if and only if $f$ is correctable (cf.\ Remark~\ref{rmk:correctablecodes}). Briefly, a disintegration is a particular kind of Bayesian inverse that we define momentarily. This provides an additional interpretation of the conditional information loss, namely as a deviation from correctability. 

\bd
\label{DD19}
Let $(Y,q)\xstoch{g} (X,p)$ be a morphism in $\FinPS$. 
Then $(X,p)\xstoch{f} (Y,q)$ is said to be a \define{disintegration} of $g$ (or $(g,q,p)$ for clarity) if and only if $g\circ f\aeequals{p}\text{id}_X$.
\ed

\blem
\label{lem:disintdeterm}
If $f$ is a disintegration of $g$, then $g$ is $q$-$a.e.$ deterministic and $f$ is a Bayesian inverse of $g$. Conversely, if $g$ is $q$-a.e.\ deterministic, then every Bayesian inverse of $g$ is a disintegration of $g$.  
\elem

\bprf
This is proved in a more abstract setting in~\cite[Section~8]{PaBayes}.
\eprf

\bt\label{DT17}
Let $(X,p)\xstoch{f} (Y,q)$ be a morphism in $\FinPS$. Then $K(f)=0$ if and only if there exists a 
map $(Y,q)\xstoch{g} (X,p)$ such that $f$ is a disintegration of $g$.
\et

\bprf
The theorem will be proved by showing the equivalent statement `$H(\ov f|q)=0$ if and only if $\ov f$ is $q$-a.e.\ deterministic for some Bayesian inverse $\ov f$ of $f$.' We therefore first justify this as being equivalent to the claim. 

First, $K(f)=0$ holds if and only if $H(\ov f)=0$ for some (and hence any) Bayesian inverse $\ov f$ of $f$ by Corollary~\ref{prop:Bayesswaps} and Lemma~\ref{reflex}. Second, the statement `there exists a $g$ such that $f$ is a disintegration of $g$' holds if and only if `there exists a $g$ such that $f$ is a Bayesian inverse of $g$ \emph{and} $g$ is $q$-a.e.\ deterministic' by Lemma~\ref{lem:disintdeterm}. However, since Bayesian inverses always exist (Theorem~\ref{thm:classicalBayestheorem}), and because Bayesian inversion is symmetric (item~\ref{item:binvsymm} in Proposition~\ref{lem:Binversionproperties}), this latter statement is equivalent to `there exists a $q$-a.e.\ deterministic Bayesian inverse $g$ of $f$.' 

Hence, suppose $f$ has a $q$-a.e.\ deterministic Bayesian inverse $g$. Then
\[
H(g|q)=\sum_{y\in Y}q_{y}H(g_{y})
=\sum_{y\in Y\setminus N_{q}}q_{y}H(g_{y})=0,
\]
since the entropy of $g_{y}$ vanishes because it is $\{0,1\}$-valued for all $y\in Y\setminus N_{q}$.  

Conversely, suppose $H(\ov f)=0$ for some Bayesian inverse $\ov f$ of $f$. Then $H(\ov f)=\sum_{y\in Y\setminus N_{q}}q_{y}H(\ov f_{y})$ is a sum of non-negative numbers that vanishes. Hence, $H(\ov f_{y})=0$ for all $y\in Y\setminus N_{q}$. But since the entropy of a probability measure on a finite set vanishes if and only if the probability measure is $\{0,1\}$-valued, $\ov f_{y}$ is $\{0,1\}$-valued for all $y\in Y\setminus N_{q}$. Hence, $\ov f$ is $q$-a.e.\ deterministic. 
\eprf

\br[Vanishing of Conditional Information Loss in Terms of Correctable Codes]\label{rmk:correctablecodes}
The vanishing of the conditional information loss is closely related to the correctability of classical codes.%
\footnote{Our references for correctable codes include~\cites{Ch06,KLABVZ02}, though our particular emphasis in terms of possibilistic maps instead of stochastic maps appears to be new. The correctability of classical codes does not require the datum of a stochastic map, but rather that of a possibilistic map.}
A \define{possibilistic map} (also called a \define{full relation}) from a finite set $X$ to a finite set $Y$ is an assignment $f$ sending $x\in X$ to a nonempty subset $f_{x}\subseteq Y$. Such a map can also be viewed as a transition kernel $X\xstoch{f}Y$ such that $f_{yx}\in\{0,1\}$ for all $x\in X$ and $y\in Y$ and for each $x\in X$ there exists a $y\in Y$ such that $f_{yx}=1$.
A \define{classical code} is a tuple $(A,X,Y,E,N)$ consisting of finite sets $A, X, Y$, an inclusion $A\xhookrightarrow{E}X$ (the \define{encoding}), and a possibilistic map $X\xstoch{N}Y$ (the \define{noise}). Such a classical code is \define{correctable} iff there exists a possibilistic map $Y\xstoch{D}A$ (the \define{recovery map}) such that $D\circ N\circ E=\id_{A}$. 

Associated with every morphism $(X,p)\xstoch{f}(Y,q)$ in $\FinPS$ is a classical code given by 
\be
\label{eq:AXYEN}
\big(A:=X\setminus N_{p},X,Y,E:=\mathrm{incl},N:=\lceil f\rceil\big),
\ee
where $X\setminus N_{p}\xhookrightarrow{\mathrm{incl}}X$ denotes the usual inclusion and where $\lceil f\rceil$ is the possibilistic map defined by 
\[
\lceil f\rceil_{yx}:=\lceil f_{yx}\rceil
:=\begin{cases}1&\mbox{ if }f_{yx}>0\\
0&\mbox{ otherwise }\end{cases}
\]
as a transition kernel, or equivalently
\[
\lceil f\rceil_{x}:=\big\{y\in Y\;:\;f_{yx}>0\big\}
\]
as a full relation. 

Now, if $K(f)=0$, then by Theorem~\ref{DT17}, there exists a $(Y,q)\xstoch{g}(X,p)$ such that $f$ is a disintegration of $(g,q,p)$. Thus, $g\circ f\aeequals{p}\id_{X}$. Since $A:=X\setminus N_{p}$, the map $g$ restricts to a deterministic map $g_{\upharpoonright}:B\to A$, where $B:=\bigcup_{x\in X}\lceil f\rceil_{x}\subseteq Y$. Since $g_{\upharpoonright}$ is deterministic, it is also possibilistic. Let $D:Y\stoch A$ be any extension of $g_{\upharpoonright}$ to a possibilistic map. This map satisfies $D\circ N\circ E=\id_{A}$ precisely because $g\circ f\aeequals{p}\id_{X}$. Thus, $(A,X,Y,E,N)$ is correctable.

Conversely, suppose $(A,X,Y,E,N)$ as in~(\ref{eq:AXYEN}) is correctable, with a possibilistic recovery map $Y\xstoch{D}A$. Then $N$ restricts to a deterministic map $N_{\upharpoonright}:B\to A$, which is, in particular, a stochastic map. Thus, set $g:Y\stoch X$ to be the stochastic map given by the composite $X\xstoch{N}A\xhookrightarrow{E}X$. Then $f$ is a disintegration of $(g,q,p)$. 

This gives a physical interpretation to the vanishing of conditional information loss. Namely, $K(f)=0$ if and only if $(A:=X\setminus N_{p},X,Y,E:=\mathrm{incl},N:=\lceil f\rceil)$ is correctable. 
\er

\section{The Markov category setting}
\label{app:markovcat}

In this appendix, we gather some definitions and results that indicate how our formalism extends to the setting of Markov categories~\cites{ChJa18,Fr19} in terms of string diagrams~\cites{Se10}. 

\bd
A \define{Markov category} is a symmetric monoidal category $(\mM,\otimes,I)$, with $\otimes$ the tensor product and $I$ the unit (associators and unitors are excluded from the notation), and where each object $X$ in $\mM$ is equipped with morphisms $!_{X}\equiv\vcenter{\hbox{%
\begin{tikzpicture}[font=\footnotesize]
\node[discarder] (d) at (0,0.15) {};
\node at (0.15,-0.15) {\scriptsize $X$};
\draw (d) to (0,-0.25);
\end{tikzpicture}}}: X\to I$ and $\Delta_{X}\equiv\vcenter{\hbox{%
\begin{tikzpicture}
\node[copier] (c) at (0,0.4) {};
\draw (c)
to[out=15,in=-90] (0.25,0.7);
\draw (c)
to[out=165,in=-90] (-0.25,0.7);
\draw (c) to (0,0.1);
\end{tikzpicture}}}: X\to X\otimes X,$ all satisfying the following conditions 
\[
\vcenter{\hbox{%
\begin{tikzpicture}[font=\footnotesize]
\node[copier] (c) at (0,0) {};
\coordinate (x1) at (-0.3,0.3);
\node[discarder] (d) at (x1) {};
\coordinate (x2) at (0.3,0.5);
\draw (c) to[out=165,in=-90] (x1);
\draw (c) to[out=15,in=-90] (x2);
\draw (c) to (0,-0.3);
\end{tikzpicture}}}
\quad=\quad
\vcenter{\hbox{%
\begin{tikzpicture}[font=\footnotesize]
\draw (0,0) to (0,0.8);
\end{tikzpicture}}}
\quad=\quad
\vcenter{\hbox{%
\begin{tikzpicture}[font=\footnotesize]
\node[copier] (c) at (0,0) {};
\coordinate (x1) at (-0.3,0.5);
\coordinate (x2) at (0.3,0.3);
\node[discarder] (d) at (x2) {};
\draw (c) to[out=165,in=-90] (x1);
\draw (c) to[out=15,in=-90] (x2);
\draw (c) to (0,-0.3);
\end{tikzpicture}}}
\qquad\qquad
\vcenter{\hbox{%
\begin{tikzpicture}[font=\footnotesize]
\node[copier] (c2) at (0,0) {};
\node[copier] (c1) at (-0.3,0.3) {};
\draw (c2) to[out=165,in=-90] (c1);
\draw (c2) to[out=15,in=-90] (0.4,0.6);
\draw (c1) to[out=165,in=-90] (-0.6,0.6);
\draw (c1) to[out=15,in=-90] (0,0.6);
\draw (c2) to (0,-0.3);
\end{tikzpicture}}}
\quad=\quad
\vcenter{\hbox{%
\begin{tikzpicture}[font=\footnotesize]
\node[copier] (c2) at (-0.5,0) {};
\node[copier] (c1) at (-0.2,0.3) {};
\draw (c2) to[out=15,in=-90] (c1);
\draw (c2) to[out=165,in=-90] (-1,0.6);
\draw (c1) to[out=165,in=-90] (-0.5,0.6);
\draw (c1) to[out=15,in=-90] (0.1,0.6);
\draw (c2) to (-0.5,-0.3);
\end{tikzpicture}}}
\qquad\qquad
\vcenter{\hbox{%
\begin{tikzpicture}[font=\small]
\node[copier] (c) at (0,0.4) {};
\draw (c)
to[out=15,in=-90] (0.25,0.65)
to[out=90,in=-90] (-0.25,1.2);
\draw (c)
to[out=165,in=-90] (-0.25,0.65)
to[out=90,in=-90] (0.25,1.2);
\draw (c) to (0,0.1);
\end{tikzpicture}}}
\quad=\quad
\vcenter{\hbox{%
\begin{tikzpicture}[font=\small]
\node[copier] (c) at (0,0.4) {};
\draw (c)
to[out=15,in=-90] (0.25,0.7);
\draw (c)
to[out=165,in=-90] (-0.25,0.7);
\draw (c) to (0,0.1);
\end{tikzpicture}}}
\]
\[
\vcenter{\hbox{%
\begin{tikzpicture}[font=\footnotesize]
\node[discarder] (d) at (0,0) {};
\draw (d) to +(0,-0.5);
\node at (0.5,-0.3) {$X\otimes Y$};
\end{tikzpicture}}}
=
\vcenter{\hbox{%
\begin{tikzpicture}[font=\footnotesize]
\node[discarder] (d) at (0,0) {};
\node[discarder] (d2) at (0.6,0) {};
\draw (d) to +(0,-0.5);
\draw (d2) to +(0,-0.5);
\node at (0.2,-0.3) {$X$};
\node at (0.8,-0.3) {$Y$};
\end{tikzpicture}}}
\qquad\quad
\vcenter{\hbox{%
\begin{tikzpicture}[font=\footnotesize]
\node[discarder] (d) at (0,0) {};
\draw (d) to +(0,-0.5);
\node at (0.2,-0.3) {$I$};
\end{tikzpicture}}}
=\;
\vcenter{\hbox{%
\begin{tikzpicture}[font=\footnotesize]
\node at (0.2,-0.05) {};
\draw [gray,dashed] (0,0) rectangle (0.45,0.65);
\end{tikzpicture}}}
\qquad\quad
\vcenter{\hbox{%
\begin{tikzpicture}[font=\footnotesize]
\node[copier] (c) at (0,0.4) {};
\draw (c)
to[out=15,in=-90] (0.25,0.7);
\draw (c)
to[out=165,in=-90] (-0.25,0.7);
\draw (c) to (0,0);
\node at (0.5,0.1) {$X\otimes Y$};
\end{tikzpicture}}}
=
\vcenter{\hbox{%
\begin{tikzpicture}[font=\footnotesize]
\node[copier] (c) at (0,0) {};
\node[copier] (c2) at (0.4,0) {};
\draw (c) to[out=15,in=-90] +(0.5,0.45);
\draw (c) to[out=165,in=-90] +(-0.4,0.45);
\draw (c) to +(0,-0.4);
\draw (c2) to[out=15,in=-90] +(0.4,0.45);
\draw (c2) to[out=165,in=-90] +(-0.5,0.45);
\draw (c2) to +(0,-0.4);
\node at (-0.2,-0.3) {$X$};
\node at (0.6,-0.3) {$Y$};
\end{tikzpicture}}}
\qquad\quad
\vcenter{\hbox{%
\begin{tikzpicture}[font=\footnotesize]
\node[copier] (c) at (0,0.4) {};
\draw (c)
to[out=15,in=-90] (0.25,0.7);
\draw (c)
to[out=165,in=-90] (-0.25,0.7);
\draw (c) to (0,0);
\node at (0.2,0.1) {$I$};
\end{tikzpicture}}}
=\;
\vcenter{\hbox{%
\begin{tikzpicture}[font=\footnotesize]
\node at (0.2,-0.05) {};
\draw [gray,dashed] (0,0) rectangle (0.45,0.75);
\end{tikzpicture}}}
\]
expressed using string diagrams. In addition, every morphism $X\xrightarrow{f}Y$ is  natural with respect to $\vcenter{\hbox{%
\begin{tikzpicture}[font=\footnotesize]
\node[discarder] (d) at (0,0.15) {};
\draw (d) to (0,-0.25);
\end{tikzpicture}}}$ in the sense that $\;\vcenter{\hbox{%
\begin{tikzpicture}[font=\footnotesize]
\node[arrow box] (c) at (0,0) {$f$};
\node[discarder] (d) at (0,0.45) {};
\draw  (c) to (d);
\draw (c) to (0,-0.45);
\end{tikzpicture}}}
\;=\;
\vcenter{\hbox{%
\begin{tikzpicture}[font=\small]
\node[discarder] (d) at (0,0) {};
\draw (d) to (0,-0.5);
\end{tikzpicture}}}$. 
\vspace{-2mm}
A \define{state} on $X$ is a morphism $I\xrightarrow{p}X$, which is drawn as $\vcenter{\hbox{%
\begin{tikzpicture}[font=\small]
\node[state] (p) at (0,0) {\footnotesize$p$};
\node at (0.15,0.35) {\scriptsize$X$};
\node (X) at (0,0.6) {};
\draw (p) to (X);
\end{tikzpicture}}}$\,.
\ed

$\FinStoch$ is a Markov category (cf.\ Section~\ref{sec:FinPS}). 
Although the definitions and results that follow are stated for stochastic maps, many hold for arbitrary Markov categories as well. 

\bd
[Definition~\ref{defn:jointdistribution} in body]
Let $(X,p)\xstoch{f} (Y,q)$ be a morphism in $\FinPS$. The \define{joint distribution} associated with $f$ is given by the following commutative diagram/string diagram equality:
\[
\left.
\xy0;/r.20pc/:
(-15,7.5)*+{\bullet}="1";
(-15,-7.5)*+{X\times Y}="2";
(15,7.5)*+{X}="0";
(15,-7.5)*+{X\times X}="3";
(-5,0)*{:=};
{\ar@{~>}"1";"2"_{\vartheta(f)}};
{\ar@{~>}"1";"0"^{p}};
{\ar@{~>}"3";"2"^{\text{id}_X\times f}};
{\ar"0";"3"^{\Delta_X}};
\endxy
\qquad\middle/\qquad
\vcenter{\hbox{%
\begin{tikzpicture}[font=\small]
\node[state] (omega) at (0,0) {\!$\vartheta(f)$\!};
\coordinate (X) at (-0.25,0.55) {};
\coordinate (Y) at (0.25,0.55) {};
\draw (omega) ++(-0.25, 0) to (X);
\draw (omega) ++(0.25, 0) to (Y);
\path[scriptstyle]
node at (-0.45,0.4) {$X$}
node at (0.45,0.4) {$Y$};
\end{tikzpicture}}}
\quad
:=
\quad
\vcenter{\hbox{%
\begin{tikzpicture}[font=\small,scale=0.90]
\node[state] (p) at (0,0) {$p$};
\node[copier] (copier) at (0,0.3) {};
\node[arrow box] (f) at (0.5,0.95) {$f$};
\coordinate (X) at (-0.5,1.5);
\coordinate (Y) at (0.5,1.5);
\draw (p) to (copier);
\draw (copier) to[out=165,in=-90] (X);
\draw (copier) to[out=15,in=-90] (f);
\draw (f) to (Y);
\path[scriptstyle]
node at (-0.7,1.45) {$X$}
node at (0.7,1.45) {$Y$};
\end{tikzpicture}}}
\right.
\]
\ed

\bn
[Extending Proposition~\ref{prop:mediatorequivalence}]
\label{prop:mediatorstringabstract}
The composable pair $(X,p)\xstoch{f} (Y,q)\xstoch{g} (Z,r)$ in $\FinPS$ is a.e.\ coalescable if and only if there exists a deterministic morphism $Z\times X\xrightarrow{h}Y$ such that
\be
\label{eq:conditionalfunctorialityzerodeviation}
\vcenter{\hbox{
\begin{tikzpicture}[font=\small,scale=0.85]
\node[state] (p) at (0.5,-0.5) {$p$};
\node[copier] (c0) at (0.5,-0.1) {};
\node[copier] (c) at (0,0.3) {};
\node[copier] (c2) at (-0.5,2.35) {};
\node[arrow box] (f) at (-0.5,0.95) {$f$};
\node[arrow box] (g) at (-0.5,1.75) {$g$};
\node[arrow box] (h) at (0.3,3.05) {\;\;$h$\;\;};
\coordinate (X) at (1.2,3.65);
\coordinate (Y1) at (0.3,3.65);
\coordinate (Y2) at (-1,3.65);
\draw (p) to (c0);
\draw (c0) to[out=15,in=-90] (X);
\draw (c) to[out=15,in=-90] (h.-60);
\draw (c0) to[out=165,in=-90] (c);
\draw (c) to[out=165,in=-90] (f);
\draw (f) to (g);
\draw (c2) to[out=15,in=-90] (h.-120);
\draw (h) to (Y1);
\draw (c2) to[out=165,in=-90] (Y2);
\draw (g) to (c2);
\end{tikzpicture}}}
\quad=\quad
\vcenter{\hbox{
\begin{tikzpicture}[font=\small]
\node[state] (p) at (0,-0.25) {$p$};
\node[copier] (c) at (0,0.3) {};
\node[copier] (c2) at (-0.5,1.55) {};
\node[arrow box] (f) at (-0.5,0.95) {$f$};
\node[arrow box] (g) at (-1,2.25) {$g$};
\coordinate (X) at (0.7,2.85);
\coordinate (Y1) at (0,2.85);
\coordinate (Y2) at (-1,2.85);
\draw (p) to (c);
\draw (c) to[out=15,in=-90] (X);
\draw (c) to[out=165,in=-90] (f);
\draw (c2) to[out=165,in=-90] (g);
\draw (g) to (Y2);
\draw (c2) to[out=15,in=-90] (Y1);
\draw (f) to (c2);
\end{tikzpicture}}}
\ee
\en

\bprf
Item~(\ref{item:mediatorstringexplicit}) of Proposition~\ref{prop:mediatorequivalence} is a direct translation of this string diagram in terms of the composition of stochastic maps between finite sets. 
\eprf

\br
\label{rmk:conditionals}
The morphism $h$ in 
Proposition~\ref{prop:mediatorstringabstract} is closely related to the abstract notion of conditionals in Markov categories~\cite[Definition~11.5]{Fr19}. Indeed, given morphisms $X\xstoch{F}Z\times Y$ and $I\xstoch{p}X$ in a Markov category, an \define{a.e.\ conditional} of $F$ given $Z$ is a morphism $Z\times X\xstoch{F|_{Z}}Y$ such that
\[
\vcenter{\hbox{%
\begin{tikzpicture}[font=\small,scale=0.85]
\node[state] (p) at (0.5,-0.6) {$p$};
\node[copier] (q) at (0.5,-0.2) {};
\node[copier] (copier) at (0,0.3) {};
\node[arrow box] (g) at (-0.5,0.95) {\;$F$\;};
\node[copier] (X) at (-0.7,1.8) {};
\coordinate (Z) at (-1.2,3.2);
\node[arrow box] (h) at (-0.1,2.6) {\;$F_{|Z}$\;};
\coordinate (YY) at (-0.1,3.2);
\coordinate (XX) at (1.0,3.2);
\node[discarder] (Y) at (-0.3,1.5) {};
\coordinate (X2) at (0.5,1.3);
\draw (q) to[out=165,in=-90] (copier);
\draw (copier) to[out=165,in=-90] (g);
\draw (copier) to[out=15,in=-90] (X2);
\draw (h.south)++(0.2,0) to[out=-90,in=90] (X2);
\draw (g.north)++(-0.2,0) to (X);
\draw (g.north)++(0.2,0) to (Y);
\draw (h.south)++(-0.2,0) to[out=-90,in=15] (X);
\draw (X) to[out=165,in=-90] (Z);
\draw (h) to (YY);
\draw (p) to (q);
\draw (q) to[out=15,in=-90] (XX);
\path[scriptstyle]
node at (-1.05,3.15) {$Z$}
node at (0.05,3.15) {$Y$}
node at (1.15,3.15) {$X$};
\end{tikzpicture}}}
\qquad
=
\qquad
\vcenter{\hbox{%
\begin{tikzpicture}[font=\small]
\node[state] (q) at (0,0) {$p$};
\node[copier] (copier) at (0,0.3) {};
\node[arrow box] (g) at (-0.5,0.95) {\;$F$\;};
\coordinate (X) at (-0.7,1.6);
\coordinate (Y) at (-0.3,1.6);
\coordinate (X2) at (0.5,1.6);
\draw (q) to (copier);
\draw (copier) to[out=165,in=-90] (g);
\draw (copier) to[out=15,in=-90] (X2);
\draw (g.north)++(-0.2,0) to (X);
\draw (g.north)++(0.2,0) to (Y);
\path[scriptstyle]
node at (-0.9,1.55) {$Z$}
node at (-0.1,1.55) {$Y$}
node at (0.7,1.55) {$X$};
\end{tikzpicture}}}
.
\]
In our case, $F=(g\times\id_{Y})\circ\Delta_{Y}\circ f$ and $h$ is the mediator. 
Therefore, a mediator is a deterministic (or at least a.e.\ deterministic) a.e.\ conditional for a specific morphism constructed from a pair of composable morphisms.
\er

\br
In string-diagram notation, Lemma~\ref{lem:conditionalmemoryadds} reads
\[
H\left(\vcenter{\hbox{
\begin{tikzpicture}[font=\footnotesize,scale=0.95]
\node[arrow box] (f) at (0,-0.3) {$f$};
\node[copier] (copier) at (0,0.3) {};
\node[arrow box] (g) at (-0.5,0.95) {$g$};
\node at (-0.7,1.45) {$Z$};
\node at (0.7,1.45) {$Y$};
\node at (-0.2,-0.75) {$X$};
\coordinate (X) at (0.5,1.6);
\coordinate (Y) at (-0.5,1.6);
\draw (0,-0.9) to (f);
\draw (f) to (copier);
\draw (copier) to[out=30,in=-90] (X);
\draw (copier) to[out=165,in=-90] (g);
\draw (g) to (Y);
\end{tikzpicture}}}
\middle|\;
\vcenter{\hbox{%
\begin{tikzpicture}[font=\small]
\node[state] (p) at (0,0) {$p$};
\node at (0.25,0.5) {\footnotesize$X$};
\node (X) at (0,0.8) {};
\draw (p) to (X);
\end{tikzpicture}}}
\right)
=
H\left(
\vcenter{\hbox{
\begin{tikzpicture}[font=\small]
\node[arrow box] (g) at (0,0) {$g$};
\coordinate (X) at (0,1.0);
\draw (0,-1.0) to (g);
\draw (g) to (X);
\node at (-0.25,0.75) {\footnotesize$Z$};
\node at (-0.25,-0.75) {\footnotesize$Y$};
\end{tikzpicture}}}
\;\middle|\;
\vcenter{\hbox{%
\begin{tikzpicture}[font=\small]
\node[state] (q) at (0,0) {$q$};
\node at (0.25,0.5) {\footnotesize$Y$};
\node (X) at (0,0.8) {};
\draw (p) to (X);
\end{tikzpicture}}}
\right)
+
H\left(
\vcenter{\hbox{
\begin{tikzpicture}[font=\small]
\node[arrow box] (f) at (0,0) {$f$};
\coordinate (X) at (0,1.0);
\draw (0,-1.0) to (f);
\draw (f) to (X);
\node at (-0.25,0.75) {\footnotesize$Y$};
\node at (-0.25,-0.75) {\footnotesize$X$};
\end{tikzpicture}}}
\;\middle|\;
\vcenter{\hbox{%
\begin{tikzpicture}[font=\small]
\node[state] (p) at (0,0) {$p$};
\node at (0.25,0.5) {\footnotesize$X$};
\node (X) at (0,0.8) {};
\draw (p) to (X);
\end{tikzpicture}}}
\right)
.
\]
\er

\bx
[Example~\ref{ex:fdetermimpliescoalescable} in body]
The mediator $Z\times X\xrightarrow{h}Y$ in this case may be given by 
\[
\vcenter{\hbox{%
\begin{tikzpicture}[font=\small]
\node[arrow box] (h) at (0,0) {\;\;$h$\;\;};
\draw (h.south)++(-0.25,0) to (-0.25,-0.7);
\draw (h.south)++(0.25,0) to (0.25,-0.7);
\draw (h) to (0,0.7);
\end{tikzpicture}}}
\quad:=\quad
\vcenter{\hbox{%
\begin{tikzpicture}[font=\small]
\node[discarder] (d) at (-0.4,0) {};
\node[arrow box] (c) at (0.4,0) {$f$};
\draw (d) to (-0.4,-0.7);
\draw  (c) to (0.4,0.7);
\draw (c) to (0.4,-0.7);
\end{tikzpicture}}}
\quad.
\]
The following string-diagrammatic calculation
\[
\vcenter{\hbox{
\begin{tikzpicture}[font=\small,scale=0.85]
\node[state] (p) at (0.5,-0.5) {$p$};
\node[copier] (c0) at (0.5,-0.1) {};
\node[copier] (c) at (0,0.3) {};
\node[copier] (c2) at (-0.5,2.35) {};
\node[arrow box] (f) at (-0.5,0.95) {$f$};
\node[arrow box] (g) at (-0.5,1.75) {$g$};
\node[arrow box] (h) at (0.6,3.05) {$f$};
\node[discarder] (d) at (0,2.95) {};
\coordinate (X) at (1.2,3.65);
\coordinate (Y1) at (0.6,3.65);
\coordinate (Y2) at (-1,3.65);
\draw (p) to (c0);
\draw (c0) to[out=15,in=-90] (X);
\draw (c) to[out=15,in=-90] (h);
\draw (c0) to[out=165,in=-90] (c);
\draw (c) to[out=165,in=-90] (f);
\draw (f) to (g);
\draw (c2) to[out=15,in=-90] (d);
\draw (h) to (Y1);
\draw (c2) to[out=165,in=-90] (Y2);
\draw (g) to (c2);
\end{tikzpicture}}}
\quad=\quad
\vcenter{\hbox{
\begin{tikzpicture}[font=\small]
\node[state] (q) at (0,-0.3) {$p$};
\node[copier] (c) at (0,0.3) {};
\node[copier] (c2) at (-0.5,0.95) {};
\node[arrow box] (g) at (0,1.65) {$f$};
\coordinate (X) at (0.7,2.95);
\node[arrow box] (g2) at (-1,1.65) {$f$};
\node[arrow box] (f) at (-1,2.4) {$g$};
\coordinate (Y1) at (-1,2.95);
\coordinate (Y2) at (0,2.95);
\draw (q) to (c);
\draw (c) to[out=15,in=-90] (X);
\draw (c) to[out=165,in=-90] (c2);
\draw (c2) to[out=15,in=-90] (g);
\draw (g2) to (f); 
\draw (f) to (Y1); 
\draw (g) to (Y2);
\draw (c2) to[out=165,in=-90] (g2);
\end{tikzpicture}}}
\quad=\quad
\vcenter{\hbox{
\begin{tikzpicture}[font=\small]
\node[state] (p) at (0,-0.3) {$p$};
\node[copier] (c) at (0,0.3) {};
\node[copier] (c2) at (-0.5,1.55) {};
\node[arrow box] (f) at (-0.5,0.95) {$f$};
\node[arrow box] (g) at (-1,2.25) {$g$};
\coordinate (X) at (0.7,2.85);
\coordinate (Y1) at (0,2.85);
\coordinate (Y2) at (-1,2.85);
\draw (p) to (c);
\draw (c) to[out=15,in=-90] (X);
\draw (c) to[out=165,in=-90] (f);
\draw (c2) to[out=165,in=-90] (g);
\draw (g) to (Y2);
\draw (c2) to[out=15,in=-90] (Y1);
\draw (f) to (c2);
\end{tikzpicture}}}
\quad,
\]
where $p$-a.e.\ determinism of $f$ was used in the second equality, 
shows that~(\ref{eq:conditionalfunctorialityzerodeviation}) holds. 
\ex

\bd
[Definition~\ref{RBIF71} in body]
\label{defn:bayesinversestring}
Let $(X,p)\xstoch{f}(Y,q)$ be a morphism in $\FinPS$. A \define{Bayesian inverse} of a morphism $(X,p)\xstoch{f}(Y,q)$ in $\FinPS$ is a morphism $(Y,q)\xstoch{\ov f}(X,p)$ such that the 
following diagram commutes/string diagram equality holds:
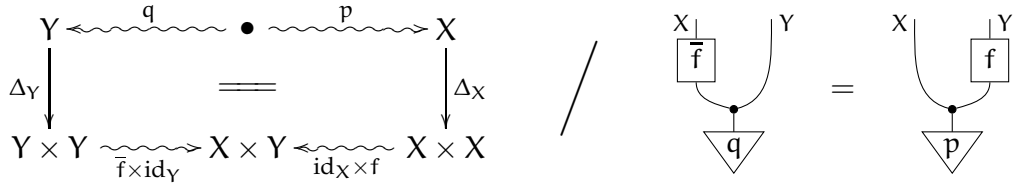
\[
\left.
\xy0;/r.25pc/:
(0,7.5)*+{\bullet}="1";
(-25,7.5)*+{Y}="Y";
(25,7.5)*+{X}="X";
(-25,-7.5)*+{Y\times Y}="YY";
(25,-7.5)*+{X\times X}="XX";
(0,-7.5)*+{X\times Y}="XY";
{\ar@{~>}"1";"X"^{p}};
{\ar@{~>}"1";"Y"_{q}};
{\ar"Y";"YY"_{\Delta_{Y}}};
{\ar"X";"XX"^{\Delta_{X}}};
{\ar@{~>}"YY";"XY"_{\ov f\times\id_{Y}}};
{\ar@{~>}"XX";"XY"^{\id_{X}\times f}};
(0,0)*{=\joinrel=\joinrel=};
\endxy
\qquad
\middle/
\qquad
\vcenter{\hbox{%
\begin{tikzpicture}[font=\small]
\node[state] (q) at (0,0) {$q$};
\node[copier] (copier) at (0,0.3) {};
\node[arrow box] (g) at (-0.5,0.95) {$\ov f$};
\coordinate (X) at (-0.5,1.5);
\coordinate (Y) at (0.5,1.5);
\draw (q) to (copier);
\draw (copier) to[out=165,in=-90] (g);
\draw (copier) to[out=15,in=-90] (Y);
\draw (g) to (X);
\path[scriptstyle]
node at (-0.7,1.45) {$X$}
node at (0.7,1.45) {$Y$};
\end{tikzpicture}}}
\;\;
=
\;\;
\vcenter{\hbox{%
\begin{tikzpicture}[font=\small]
\node[state] (p) at (0,0) {$p$};
\node[copier] (copier) at (0,0.3) {};
\node[arrow box] (f) at (0.5,0.95) {$f$};
\coordinate (X) at (-0.5,1.5);
\coordinate (Y) at (0.5,1.5);
\draw (p) to (copier);
\draw (copier) to[out=165,in=-90] (X);
\draw (copier) to[out=15,in=-90] (f);
\draw (f) to (Y);
\path[scriptstyle]
node at (-0.7,1.45) {$X$}
node at (0.7,1.45) {$Y$};
\end{tikzpicture}}}
\right.
\]
\ed

\bprf[Alternative proof of Propotion~\ref{BICP979}]
A more abstract proof of Propotion~\ref{BICP979} that is valid in an arbitrary classical Markov category can be given as follows:  
\begingroup
\allowdisplaybreaks
\begin{align*}
\!\!\!
\vcenter{\hbox{
\begin{tikzpicture}[font=\small]
\node[state] (p) at (0.5,-0.5) {$r$};
\node[copier] (c0) at (0.5,-0.1) {};
\node[copier] (c) at (0,0.3) {};
\node[copier] (c2) at (-0.5,2.35) {};
\node[arrow box] (f) at (-0.5,0.95) {$\ov g$};
\node[arrow box] (g) at (-0.5,1.75) {$\ov f$};
\node[arrow box] (h) at (0.3,3.55) {\;\;$h$\;\;};
\coordinate (hL) at (0.05,2.75);
\coordinate (hR) at (0.55,2.75);
\coordinate (X) at (1.2,4.15);
\coordinate (Y1) at (0.3,4.15);
\coordinate (Y2) at (-1,4.15);
\draw (p) to (c0);
\draw (c0) to[out=15,in=-90] (X);
\draw (c) to[out=15,in=-90] (hR);
\draw (hR) to[out=90,in=-90] (h.-130);
\draw (c0) to[out=165,in=-90] (c);
\draw (c) to[out=165,in=-90] (f);
\draw (f) to (g);
\draw (c2) to[out=15,in=-90] (hL);
\draw (hL) to[out=90,in=-90] (h.-50);
\draw (h) to (Y1);
\draw (c2) to[out=165,in=-90] (Y2);
\draw (g) to (c2);
\end{tikzpicture}}}
\!\!\!\!\!&=\!\!
\vcenter{\hbox{
\begin{tikzpicture}[font=\small]
\node[state] (p) at (0.3,-0.5) {$r$};
\node[copier] (c0) at (0.3,-0.1) {};
\node[copier] (c) at (1.1,2.35) {}; 
\node[copier] (c2) at (-0.5,2.35) {};
\node[arrow box] (f) at (-0.5,0.95) {$\ov g$};
\node[arrow box] (g) at (-0.5,1.75) {$\ov f$};
\node[arrow box] (h) at (0.3,3.55) {\;\;$h$\;\;};
\coordinate (hL) at (0.05,2.75);
\coordinate (hR) at (0.55,2.75);
\coordinate (X) at (1.6,4.15);
\coordinate (Y1) at (0.3,4.15);
\coordinate (Y2) at (-1,4.15);
\draw (p) to (c0);
\draw (c0) to[out=15,in=-90] (c);
\draw (c) to[out=165,in=-90] (hR);
\draw (hR) to[out=90,in=-90] (h.-130);
\draw (c0) to[out=165,in=-90] (f);
\draw (c) to[out=15,in=-90] (X);
\draw (f) to (g);
\draw (c2) to[out=15,in=-90] (hL);
\draw (hL) to[out=90,in=-90] (h.-50);
\draw (h) to (Y1);
\draw (c2) to[out=165,in=-90] (Y2);
\draw (g) to (c2);
\end{tikzpicture}}}
\!\!\!=\!\!\!
\vcenter{\hbox{
\begin{tikzpicture}[font=\small]
\node[state] (p) at (0.3,-0.5) {$p$};
\node[copier] (c0) at (0.3,-0.1) {};
\node[copier] (c) at (1.1,2.35) {}; 
\node[copier] (c2) at (-0.5,2.35) {};
\node[arrow box] (f) at (1.1,0.95) {$f$};
\node[arrow box] (g) at (1.1,1.75) {$g$};
\node[arrow box] (h) at (0.3,3.55) {\;\;$h$\;\;};
\coordinate (hL) at (0.05,2.75);
\coordinate (hR) at (0.55,2.75);
\coordinate (X) at (1.6,4.15);
\coordinate (Y1) at (0.3,4.15);
\coordinate (Y2) at (-1,4.15);
\draw (p) to (c0);
\draw (c0) to[out=165,in=-90] (c2);
\draw (c) to[out=165,in=-90] (hR);
\draw (hR) to[out=90,in=-90] (h.-130);
\draw (c0) to[out=15,in=-90] (f);
\draw (c) to[out=15,in=-90] (X);
\draw (f) to (g);
\draw (c2) to[out=15,in=-90] (hL);
\draw (hL) to[out=90,in=-90] (h.-50);
\draw (h) to (Y1);
\draw (c2) to[out=165,in=-90] (Y2);
\draw (g) to (c);
\end{tikzpicture}}}
\!=\!
\vcenter{\hbox{
\begin{tikzpicture}[font=\small]
\node[state] (p) at (0.3,-1.0) {$p$};
\node[copier] (c0) at (0.3,-0.7) {};
\coordinate (R1) at (0.8,-0.3);
\coordinate (L1) at (-0.2,-0.3);
\coordinate (L1b) at (-0.5,0.7);
\node[copier] (c) at (1.1,2.35) {}; 
\node[copier] (c2) at (-0.5,2.35) {};
\node[arrow box] (f) at (1.1,0.95) {$f$};
\node[arrow box] (g) at (1.1,1.75) {$g$};
\node[arrow box] (h) at (0.3,3.95) {\;\;$h$\;\;};
\coordinate (hL) at (0.05,2.75);
\coordinate (hR) at (0.55,2.75);
\coordinate (R2) at (1.6,2.75);
\coordinate (L2) at (-1,2.75);
\coordinate (X) at (1.6,4.55);
\coordinate (Y1) at (0.3,4.55);
\coordinate (Y2) at (-1,4.55);
\draw (p) to (c0);
\draw (c0) to[out=165,in=-90] (L1);
\draw (L1) to[out=90,in=-90] (f);
\draw (c0) to[out=15,in=-90] (R1);
\draw (R1) to[out=90,in=-90] (L1b);
\draw (L1b) to (c2);
\draw (c) to[out=165,in=-90] (hR);
\draw (hR) to[out=90,in=-90] (X);
\draw (c) to[out=15,in=-90] (R2);
\draw (R2) to[out=90,in=-90] (h.-130);
\draw (f) to (g);
\draw (c2) to[out=15,in=-90] (hL);
\draw (L2) to[out=90,in=-90] (h.-50);
\draw (hL) to[out=90,in=-90] (Y2);
\draw (h) to (Y1);
\draw (c2) to[out=165,in=-90] (L2);
\draw (g) to (c);
\end{tikzpicture}}}
=
\vcenter{\hbox{
\begin{tikzpicture}[font=\small]
\node[state] (p) at (0.3,-0.5) {$p$};
\node[copier] (c0) at (0.3,-0.1) {};
\node[copier] (c) at (1.0,0.3) {};
\node[copier] (c2) at (-0.5,2.35) {};
\node[arrow box] (f) at (-0.5,0.95) {$f$};
\node[arrow box] (g) at (-0.5,1.75) {$g$};
\node[arrow box] (h) at (0.3,3.05) {\;\;$h$\;\;};
\coordinate (X) at (1.2,3.45);
\coordinate (Y1) at (0.3,3.45);
\coordinate (Y2) at (-1,3.45);
\coordinate (11) at (-1,4.2);
\coordinate (21) at (0.3,4.2);
\coordinate (31) at (1.2,4.2);
\coordinate (12) at (-1,4.8);
\coordinate (22) at (0.3,4.8);
\coordinate (32) at (1.2,4.8);
\coordinate (13) at (-1,5.4);
\coordinate (23) at (0.3,5.4);
\coordinate (33) at (1.2,5.4);
\draw (p) to (c0);
\draw (c0) to[out=165,in=-90] (f);
\draw (c) to[out=165,in=-90] (h.-60);
\draw (c0) to[out=15,in=-90] (c);
\draw (c) to[out=15,in=-90] (X);
\draw (f) to (g);
\draw (c2) to[out=15,in=-90] (h.-120);
\draw (h) to (Y1);
\draw (c2) to[out=165,in=-90] (Y2);
\draw (g) to (c2);
\draw (Y2) to[out=90,in=-90] (21);
\draw (Y1) to[out=90,in=-90] (11);
\draw (X) to[out=90,in=-90] (31);
\draw (11) to[out=90,in=-90] (12);
\draw (21) to[out=90,in=-90] (32);
\draw (31) to[out=90,in=-90] (22);
\draw (12) to[out=90,in=-90] (23);
\draw (22) to[out=90,in=-90] (13);
\draw (32) to[out=90,in=-90] (33);
\end{tikzpicture}}}
\\
&=
\vcenter{\hbox{
\begin{tikzpicture}[font=\small]
\node[state] (p) at (0,-0.1) {$p$};
\node[copier] (c) at (0,0.3) {};
\node[copier] (c2) at (-0.5,1.55) {};
\node[arrow box] (f) at (-0.5,0.95) {$f$};
\node[arrow box] (g) at (-1,2.25) {$g$};
\coordinate (X) at (0.7,2.55);
\coordinate (Y1) at (0,2.55);
\coordinate (Y2) at (-1,2.55);
\coordinate (11) at (-1,3.2);
\coordinate (21) at (0,3.2);
\coordinate (31) at (0.7,3.2);
\coordinate (12) at (-1,3.8);
\coordinate (22) at (0,3.8);
\coordinate (32) at (0.7,3.8);
\coordinate (13) at (-1,4.4);
\coordinate (23) at (0,4.4);
\coordinate (33) at (0.7,4.4);
\draw (p) to (c);
\draw (c) to[out=15,in=-90] (X);
\draw (c) to[out=165,in=-90] (f);
\draw (c2) to[out=165,in=-90] (g);
\draw (g) to (Y2);
\draw (c2) to[out=15,in=-90] (Y1);
\draw (f) to (c2);
\draw (Y2) to[out=90,in=-90] (21);
\draw (Y1) to[out=90,in=-90] (11);
\draw (X) to[out=90,in=-90] (31);
\draw (11) to[out=90,in=-90] (12);
\draw (21) to[out=90,in=-90] (32);
\draw (31) to[out=90,in=-90] (22);
\draw (12) to[out=90,in=-90] (23);
\draw (22) to[out=90,in=-90] (13);
\draw (32) to[out=90,in=-90] (33);
\end{tikzpicture}}}
\!=\!
\vcenter{\hbox{
\begin{tikzpicture}[font=\small]
\node[state] (p) at (0,-0.1) {$r$};
\node[copier] (c) at (0,0.3) {};
\node[copier] (c2) at (0.5,1.55) {};
\node[arrow box] (f) at (0.5,0.95) {$\ov g$};
\node[arrow box] (g) at (1,2.25) {$\ov f$};
\coordinate (X) at (1.0,2.55);
\coordinate (Y1) at (0,2.55);
\coordinate (Y2) at (-0.7,2.55);
\coordinate (11) at (-0.7,3.2);
\coordinate (21) at (0,3.2);
\coordinate (31) at (1.0,3.2);
\coordinate (12) at (-0.7,3.8);
\coordinate (22) at (0,3.8);
\coordinate (32) at (1.0,3.8);
\coordinate (13) at (-0.7,4.4);
\coordinate (23) at (0,4.4);
\coordinate (33) at (1.0,4.4);
\draw (p) to (c);
\draw (c) to[out=165,in=-90] (Y2);
\draw (c) to[out=15,in=-90] (f);
\draw (c2) to[out=15,in=-90] (g);
\draw (g) to (X);
\draw (c2) to[out=165,in=-90] (Y1);
\draw (f) to (c2);
\draw (Y2) to[out=90,in=-90] (21);
\draw (Y1) to[out=90,in=-90] (11);
\draw (X) to[out=90,in=-90] (31);
\draw (11) to[out=90,in=-90] (12);
\draw (21) to[out=90,in=-90] (32);
\draw (31) to[out=90,in=-90] (22);
\draw (12) to[out=90,in=-90] (23);
\draw (22) to[out=90,in=-90] (13);
\draw (32) to[out=90,in=-90] (33);
\end{tikzpicture}}}
\;=\;
\vcenter{\hbox{
\begin{tikzpicture}[font=\small]
\node[state] (q) at (0,-0.3) {$r$};
\node[copier] (c) at (0,0.1) {};
\coordinate (aL) at (-0.5,0.55);
\coordinate (aR) at (0.5,0.55);
\coordinate (aR2) at (0.7,1.35);
\coordinate (bL) at (-1,4.15);
\coordinate (bR) at (0,4.15);
\node[copier] (c2) at (-0.5,2.15) {};
\node[arrow box] (g) at (-0.5,1.55) {$\ov g$};
\node[arrow box] (f) at (0,2.85) {$\ov f$};
\coordinate (X) at (0.7,4.15);
\coordinate (Y1) at (0,3.35);
\coordinate (Y2) at (-1,3.35);
\draw (q) to (c);
\draw (c) to[out=15,in=-90] (aR);
\draw (aL) to[out=90,in=-90] (aR2);
\draw (aR2) to (X);
\draw (c) to[out=165,in=-90] (aL);
\draw (aR) to[out=90,in=-90] (g);
\draw (c2) to[out=15,in=-90] (f);
\draw (f) to (Y1);
\draw (Y1) to[out=90,in=-90] (bL);
\draw (Y2) to[out=90,in=-90] (bR);
\draw (c2) to[out=165,in=-90] (Y2);
\draw (g) to (c2);
\end{tikzpicture}}}
\;\;=\;
\vcenter{\hbox{
\begin{tikzpicture}[font=\small]
\node[state] (p) at (0,-0.1) {$r$};
\node[copier] (c) at (0,0.3) {};
\node[copier] (c2) at (-0.5,1.55) {};
\node[arrow box] (f) at (-0.5,0.95) {$\ov g$};
\node[arrow box] (g) at (-1,2.25) {$\ov f$};
\coordinate (X) at (0.7,2.85);
\coordinate (Y1) at (0,2.85);
\coordinate (Y2) at (-1,2.85);
\draw (p) to (c);
\draw (c) to[out=15,in=-90] (X);
\draw (c) to[out=165,in=-90] (f);
\draw (c2) to[out=165,in=-90] (g);
\draw (g) to (Y2);
\draw (c2) to[out=15,in=-90] (Y1);
\draw (f) to (c2);
\end{tikzpicture}}}
.\qedhere
\end{align*}
\endgroup
\eprf

\bd
[Definition~\ref{defn:bloomshriek} in body]
Given a stochastic map $X\xstoch{f}Y$, the \define{bloom} $X\xstoch{\bloom_{f}}X\times Y$ and \define{shriek} $X\times Y\xrightarrow{!_{f}}X$ of $f$ are given by 
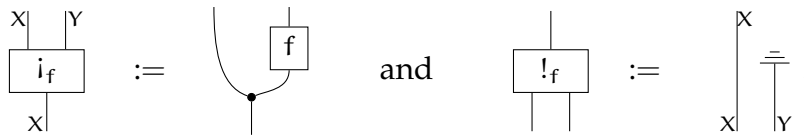
\[
\vcenter{\hbox{
\begin{tikzpicture}[font=\small]
\node[arrow box] (p) at (0,0) {\;\;$\bloom_{f}$\;\;};
\coordinate (X) at (-0.25,0.8);
\coordinate (Y) at (0.25,0.8);
\draw (0,-0.8) to (p);
\draw (p.north)++(-0.25,0) to (X);
\draw (p.north)++(0.25,0) to (Y);
\node at (-0.15,-0.70) {\scriptsize $X$};
\node at (-0.375,0.70) {\scriptsize $X$};
\node at (0.375,0.70) {\scriptsize $Y$};
\end{tikzpicture}}}
\quad
:=
\quad
\vcenter{\hbox{
\begin{tikzpicture}[font=\small,xscale=-1]
\coordinate (p) at (0,-0.2);
\node[copier] (copier) at (0,0.3) {};
\node[arrow box] (f) at (-0.5,0.95) {$f$};
\coordinate (X) at (0.5,1.5);
\coordinate (Y) at (-0.5,1.5);
\draw (p) to (copier);
\draw (copier) to[out=15,in=-90] (X);
\draw (copier) to[out=165,in=-90] (f);
\draw (f) to (Y);
\end{tikzpicture}}}
\qquad
\text{ and }
\qquad
\vcenter{\hbox{
\begin{tikzpicture}[font=\small]
\node[arrow box] (p) at (0,0) {\;\;$!_{f}$\;\;};
\coordinate (X) at (-0.25,-0.8);
\coordinate (Y) at (0.25,-0.8);
\draw (0,0.8) to (p);
\draw (p.south)++(-0.25,0) to (X);
\draw (p.south)++(0.25,0) to (Y);
\end{tikzpicture}}}
\quad
:=
\quad
\vcenter{\hbox{
\begin{tikzpicture}[font=\small]
\coordinate (X) at (-0.25,-0.8);
\coordinate (X2) at (-0.25,0.8);
\coordinate (Y) at (0.25,-0.8);
\coordinate (d) at (0.25,0.1);
\node[discarder] at (d) {}; 
\draw (X2) to (X);
\draw (d) to (Y);
\node at (-0.15,0.70) {\scriptsize $X$};
\node at (-0.375,-0.70) {\scriptsize $X$};
\node at (0.375,-0.70) {\scriptsize $Y$};
\end{tikzpicture}}}
.
\]
\ed

\bprf
[Diagrammatic proof of Propostion~\ref{prop:bloomshriekfactorization}]
{\color{white}{you found me!}}
\begin{enumerate}[i)]
\itemsep0pt
\item
First, 
\[
\vcenter{\hbox{
\begin{tikzpicture}[font=\small]
\node[arrow box] (bloom) at (0,0) {\;\;$\bloom_{f}$\;\;};
\coordinate (X) at (0,2.0);
\node[arrow box] (shriek) at (0,1.1) {\;\;$\shriek_{f}$\;\;};
\draw (0,-0.8) to (bloom);
\draw (bloom.north)++(-0.25,0) to ([xshift=-0.25cm]shriek.south);
\draw (bloom.north)++(0.25,0) to ([xshift=0.25cm]shriek.south);
\draw (shriek) to (X);
\path[scriptstyle]
node at (-0.2,-0.6) {$X$}
node at (-0.2,1.8) {$X$}
node at (-0.45,0.55) {$X$}
node at (0.45,0.55) {$Y$};
\end{tikzpicture}}}
\quad=\quad
\vcenter{\hbox{
\begin{tikzpicture}[font=\small,xscale=-1]
\coordinate (p) at (0,-0.3);
\node[copier] (copier) at (0,0.3) {};
\node[arrow box] (f) at (-0.5,0.95) {$f$};
\coordinate (X) at (0.5,1.7);
\node[discarder] (Y) at (-0.5,1.5) {};
\draw (p) to (copier);
\draw (copier) to[out=15,in=-90] (X);
\draw (copier) to[out=165,in=-90] (f);
\draw (f) to (Y);
\path[scriptstyle]
node at (0.2,-0.2) {$X$}
node at (0.7,1.5) {$X$};
\end{tikzpicture}}}
\quad=\quad
\vcenter{\hbox{
\begin{tikzpicture}[font=\small,xscale=-1]
\coordinate (p) at (0,0.3);
\coordinate (X) at (0,1.6);
\draw (p) to (X);
\end{tikzpicture}}}
\qquad\text{ and }\qquad
\vcenter{\hbox{%
\begin{tikzpicture}[font=\small]
\node[state] (omega) at (0,0) {$\!\vartheta(f)\!$};
\coordinate (X) at (-0.5,1.15) {};
\coordinate (Y) at (0.5,1.15) {};
\draw (omega) ++(-0.5, 0) to (X);
\draw (omega) ++(0.5, 0) to (Y);
\path[scriptstyle]
node at (-0.7,1.0) {$X$}
node at (0.7,1.0) {$Y$};
\end{tikzpicture}}}
\quad
=
\quad
\vcenter{\hbox{%
\begin{tikzpicture}[font=\small]
\node[state] (p) at (0,0) {$p$};
\node[copier] (copier) at (0,0.3) {};
\node[arrow box] (f) at (0.5,0.95) {$f$};
\coordinate (X) at (-0.5,1.5);
\coordinate (Y) at (0.5,1.5);
\draw (p) to (copier);
\draw (copier) to[out=165,in=-90] (X);
\draw (copier) to[out=15,in=-90] (f);
\draw (f) to (Y);
\path[scriptstyle]
node at (-0.7,1.45) {$X$}
node at (0.7,1.45) {$Y$};
\end{tikzpicture}}},
\]
where the second equality holds by the very definition of the joint distribution $\vartheta(f)$. 
\item
Secondly, 
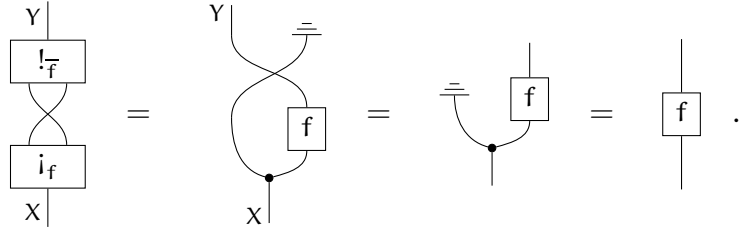
\[
\vcenter{\hbox{
\begin{tikzpicture}[font=\small]
\node[arrow box] (bloom) at (0,0) {\;\;$\bloom_{f}$\;\;};
\coordinate (Y) at (0,2.2);
\node[arrow box] (shriek) at (0,1.4) {\;\;$\shriek_{\ov f}$\;\;};
\draw (0,-0.8) to (bloom);
\draw (bloom.north)++(-0.25,0) to[out=90,in=-90] ([xshift=0.25cm]shriek.south);
\draw (bloom.north)++(0.25,0) to[out=90,in=-90] ([xshift=-0.25cm]shriek.south);
\draw (shriek) to (Y);
\path[scriptstyle]
node at (-0.2,-0.6) {$X$}
node at (-0.2,2.0) {$Y$};
\end{tikzpicture}}}
\quad=\quad
\vcenter{\hbox{
\begin{tikzpicture}[font=\small]
\coordinate (p) at (0,-0.3);
\node[copier] (copier) at (0,0.3) {};
\node[arrow box] (f) at (0.5,0.95) {$f$};
\coordinate (X) at (-0.5,1.1);
\coordinate (Y) at (-0.5,2.2) {};
\coordinate (Yend) at (-0.5,2.6) {};
\node[discarder] (X2) at (0.5,2.2) {};
\draw (p) to (copier);
\draw (copier) to[out=165,in=-90] (X);
\draw (copier) to[out=15,in=-90] (f);
\draw (f) to[out=90,in=-90] (Y);
\draw (X) to[out=90,in=-90] (X2);
\draw (Y) to (Yend);
\path[scriptstyle]
node at (-0.2,-0.2) {$X$}
node at (-0.7,2.5) {$Y$};
\end{tikzpicture}}}
\quad=\quad
\vcenter{\hbox{
\begin{tikzpicture}[font=\small]
\coordinate (p) at (0,-0.2);
\node[copier] (copier) at (0,0.3) {};
\node[arrow box] (f) at (0.5,0.95) {$f$};
\node[discarder] (X) at (-0.5,1.0) {};
\coordinate (Y) at (0.5,1.7);
\draw (p) to (copier);
\draw (copier) to[out=165,in=-90] (X);
\draw (copier) to[out=15,in=-90] (f);
\draw (f) to (Y);
\end{tikzpicture}}}
\quad=\quad
\vcenter{\hbox{
\begin{tikzpicture}[font=\small]
\node[arrow box] (p) at (0,0) {$f$};
\coordinate (X) at (0,1.0);
\draw (0,-1.0) to (p);
\draw (p) to (X);
\end{tikzpicture}}}
\quad.
\]
\item
Finally, set the mediator function $Y\times X\xrightarrow{h}X\times Y$ to be the swap map. Then
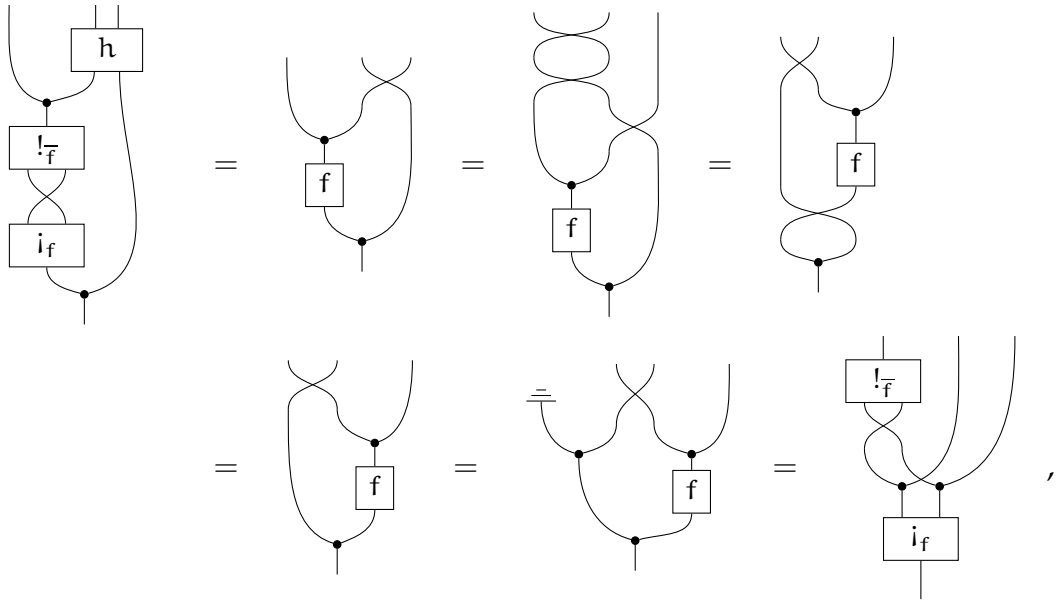
\begingroup
\allowdisplaybreaks
\begin{align*}
\vcenter{\hbox{
\begin{tikzpicture}[font=\small]
\node[copier] (c) at (0,-0.2) {};
\node[copier] (c2) at (-0.5,2.35) {};
\node[arrow box] (f) at (-0.5,0.45) {\;\;$\bloom_{f}$\;\;};
\node[arrow box] (g) at (-0.5,1.75) {\;\;$\shriek_{\ov f}$\;\;};
\node[arrow box] (h) at (0.3,3.05) {\;\;$h$\;\;};
\coordinate (Y1) at (0.45,3.65);
\coordinate (X) at (0.15,3.65);
\coordinate (Y2) at (-1,3.65);
\draw (0,-0.6) to (c);
\draw (c) to[out=15,in=-90] (h.-60);
\draw (c) to[out=165,in=-90] (f);
\draw (c2) to[out=15,in=-90] (h.-120);
\draw (h.north)++(0.15,0) to (Y1);
\draw (h.north)++(-0.15,0) to (X);
\draw (c2) to[out=165,in=-90] (Y2);
\draw (g) to (c2);
\draw (f.north)++(-0.25,0) to[out=90,in=-90] ([xshift=0.25cm]g.south);
\draw (f.north)++(0.25,0) to[out=90,in=-90] ([xshift=-0.25cm]g.south);
\end{tikzpicture}}}
\quad&=\quad
\vcenter{\hbox{
\begin{tikzpicture}[font=\small]
\node[copier] (c) at (0,-0.2) {};
\node[copier] (c2) at (-0.5,1.15) {};
\node[arrow box] (f) at (-0.5,0.55) {$f$};
\coordinate (Y1) at (0.65,2.25);
\coordinate (X2) at (0.65,1.6);
\coordinate (Y) at (0,1.6);
\coordinate (X) at (0,2.25);
\coordinate (Y2) at (-1,2.25);
\draw (0,-0.6) to (c);
\draw (c) to[out=15,in=-90] (X2);
\draw (c) to[out=165,in=-90] (f);
\draw (c2) to[out=15,in=-90] (Y);
\draw (c2) to[out=165,in=-90] (Y2);
\draw (f) to (c2); 
\draw (X2) to[out=90,in=-90] (X);
\draw (Y) to[out=90,in=-90] (Y1);
\end{tikzpicture}}}
\quad=\quad
\vcenter{\hbox{
\begin{tikzpicture}[font=\small]
\node[copier] (c) at (0,-0.2) {};
\node[copier] (c2) at (-0.5,1.15) {};
\node[arrow box] (f) at (-0.5,0.55) {$f$};
\coordinate (Y1) at (0.65,2.25);
\coordinate (X2) at (0.65,1.6);
\coordinate (Y) at (0,1.6);
\coordinate (X) at (0,2.25);
\coordinate (Y2) at (-1,2.25);
\coordinate (a) at (-1,2.85);
\coordinate (b) at (0,2.85);
\coordinate (L) at (-1,3.45);
\coordinate (M) at (0,3.45);
\coordinate (R) at (0.65,3.45);
\draw (0,-0.6) to (c);
\draw (c) to[out=15,in=-90] (X2);
\draw (c) to[out=165,in=-90] (f);
\draw (c2) to[out=15,in=-90] (Y);
\draw (c2) to[out=165,in=-90] (Y2);
\draw (f) to (c2); 
\draw (X2) to[out=90,in=-90] (X);
\draw (Y) to[out=90,in=-90] (Y1);
\draw (Y1) to (R);
\draw (Y2) to[out=90,in=-90] (b);
\draw (X) to[out=90,in=-90] (a);
\draw (a) to[out=90,in=-90] (M);
\draw (b) to[out=90,in=-90] (L);
\end{tikzpicture}}}
\quad=\quad
\vcenter{\hbox{
\begin{tikzpicture}[font=\small]
\coordinate (Xb) at (0,0);
\node[copier] (c) at (0,0.4) {};
\coordinate (X1) at (-0.5,0.7) {};
\coordinate (X2) at (0.5,0.7) {};
\node[arrow box] (f) at (0.5,1.7) {$f$};
\coordinate (X3) at (-0.5,1.4);
\coordinate (X4) at (-0.5,2.7);
\node[copier] (c2) at (0.5,2.4) {};
\coordinate (Ym) at (0,2.7);
\coordinate (Y1) at (-0.5,3.4);
\coordinate (Y2) at (1.0,3.4);
\coordinate (X5) at (0,3.4);
\draw (Xb) to (c);
\draw (c) to[out=165,in=-90] (X1);
\draw (c) to[out=15,in=-90] (X2);
\draw (X2) to[out=90,in=-90] (X3);
\draw (X1) to[out=90,in=-90] (f);
\draw (X3) to (X4);
\draw (f) to (c2);
\draw (c2) to[out=165,in=-90] (Ym);
\draw (c2) to[out=15,in=-90] (Y2);
\draw (Ym) to[out=90,in=-90] (Y1);
\draw (X4) to[out=90,in=-90] (X5);
\end{tikzpicture}}}
\\
&=\quad
\vcenter{\hbox{
\begin{tikzpicture}[font=\small,xscale=-1]
\node[copier] (c) at (0,-0.2) {};
\node[copier] (c2) at (-0.5,1.15) {};
\node[arrow box] (f) at (-0.5,0.55) {$f$};
\coordinate (Y1) at (0.65,2.25);
\coordinate (X2) at (0.65,1.6);
\coordinate (Y) at (0,1.6);
\coordinate (X) at (0,2.25);
\coordinate (Y2) at (-1,2.25);
\draw (0,-0.6) to (c);
\draw (c) to[out=15,in=-90] (X2);
\draw (c) to[out=165,in=-90] (f);
\draw (c2) to[out=15,in=-90] (Y);
\draw (c2) to[out=165,in=-90] (Y2);
\draw (f) to (c2); 
\draw (X2) to[out=90,in=-90] (X);
\draw (Y) to[out=90,in=-90] (Y1);
\end{tikzpicture}}}
\quad=\quad
\vcenter{\hbox{
\begin{tikzpicture}[font=\small]
\coordinate (p) at (-0.25,-0.1);
\node[copier] (c) at (-0.25,0.3) {};
\node[copier] (c2) at (0.5,1.45) {};
\node[arrow box] (f) at (0.5,0.95) {$f$};
\coordinate (X) at (0,2.65);
\node[copier] (c3) at (-1,1.45) {};
\coordinate (a) at (-0.5,1.85);
\coordinate (b) at (0,1.85);
\coordinate (Y2) at (-0.5,2.65);
\coordinate (Y3) at (1,2.65);
\node[discarder] (d) at (-1.5,2.15) {};
\draw (p) to (c);
\draw (c) to[out=165,in=-90] (c3);
\draw (c) to[out=15,in=-90] (f);
\draw (c2) to[out=15,in=-90] (Y3);
\draw (c2) to[out=165,in=-90] (b);
\draw (c3) to[out=15,in=-90] (a);
\draw (c3) to[out=165,in=-90] (d);
\draw (f) to (c2);
\draw (a) to[out=90,in=-90] (X);
\draw (b) to[out=90,in=-90] (Y2);
\end{tikzpicture}}}
\quad=\quad
\vcenter{\hbox{
\begin{tikzpicture}[font=\small]
\node[arrow box] (p) at (0,0) {\;\;$\bloom_{f}$\;\;};
\node[copier] (c1) at (-0.25,0.7) {};
\node[copier] (c2) at (0.25,0.7) {};
\coordinate (a) at (-0.75,1.2);
\coordinate (b) at (-0.25,1.2);
\node[arrow box] (s) at (-0.5,2.1) {\;\;$\shriek_{\ov f}$\;\;};
\coordinate (Y1) at (-0.5,2.7);
\coordinate (X1) at (0.5,2.7);
\coordinate (Y2) at (1.25,2.7);
\draw (0,-0.8) to (p);
\draw (p.north)++(-0.25,0) to (c1);
\draw (p.north)++(0.25,0) to (c2);
\draw (c1) to[out=165,in=-90] (a);
\draw (c1) to[out=15,in=-90] (X1);
\draw (c2) to[out=165,in=-90] (b);
\draw (c2) to[out=15,in=-90] (Y2);
\draw (a) to[out=90,in=-90] ([xshift=0.25cm]s.south);
\draw (b) to[out=90,in=-90] ([xshift=-0.25cm]s.south);
\draw (s) to (Y1);
\end{tikzpicture}}}
\quad,
\end{align*}
\endgroup
which proves that the pair $(X,p)\xstoch{\bloom_{f}}(X\times Y,\vartheta(f))\xrightarrow{!_{\ov f}\circ\gamma}(Y,q)$ is coalescable. \qedhere
\end{enumerate}
\eprf

\vspace{3mm}
\noindent
{\large \textbf{Acknowledgements.}}
We thank Tobias Fritz, Azeem ul~Hassan, Francois Jacopin, Jiaxin Qiao, and Alex Atsushi Takeda for discussions. 
This research has received funding from the European Research Council (ERC) under the European Union's Horizon 2020 research and innovation program (QUASIFT grant agreement 677368).

\addcontentsline{toc}{section}{\numberline{}Bibliography}
\bibliographystyle{plain}
\bibliography{conditionalBayes}

\Addresses

\end{document}